\documentclass[a4paper,11pt,headings=big,DIV=12]{scrartcl}
\pdfoutput=1
\usepackage{amsmath,amssymb,mathtools}
\usepackage[table]{xcolor}
\usepackage{tcolorbox}
\definecolor{mygray}{gray}{0.2}
 \addtokomafont{disposition}{\rmfamily\boldmath}
  \usepackage{bbold}
   \usepackage{environ}
 \renewcommand*{\arraystretch}{2.0}

\NewEnviron{thincases}{\scalebox{0.5}[1]{$\displaystyle
\left\{\scalebox{2}[1]{\setlength{\arraycolsep}{6pt}
$\displaystyle\begin{array}{ll}
\BODY
\end{array}$}\right.$}}

\definecolor{mypink1}{rgb}{0.9, 0.2, 0.6}
\usepackage{graphicx}
\addtokomafont{disposition}{\rmfamily\boldmath}
\usepackage[utf8]{inputenc}
\usepackage{enumerate}
\usepackage{slashed}
\usepackage{cite}
\usepackage{comment}
\usepackage{multirow}
\usepackage{acronym}
 \usepackage{bbold}
 \usepackage{jheppub}
\allowdisplaybreaks
\newcommand{\com}[1]{
{#1}}

\definecolor{mypink1}{rgb}{0.9, 0.2, 0.6}
\usepackage{graphicx}
\addtokomafont{disposition}{\rmfamily\boldmath}
\usepackage[utf8]{inputenc}
\usepackage{enumerate}
\usepackage{slashed}
\usepackage{cite}
\usepackage{comment}
\usepackage{multirow}
\usepackage{acronym}
 \usepackage{bbold}
\allowdisplaybreaks
\newcommand{\FC}{\bar c_\sig}
\newcommand{\xx}{z}

\numberwithin{equation}{section}

\newcommand{\cw}{\omega}

\newcommand{\vev}[1]{\langle #1 \rangle}

\newcommand{\matel}[3]{\langle #1|#2|#3\rangle}

\newcommand{\al}{\alpha}
\newcommand{\be}{\beta}
\newcommand{\ga}{\gamma}
\newcommand{\Ga}{\Gamma}
\newcommand{\de}{\delta}
\newcommand{\De}{\Delta}
\newcommand{\la}{\lambda}

\newcommand{\eps}{\epsilon}

\newcommand{\sig}{ \sigma}
\newcommand{\Sig}{ \Sigma}
\newcommand{\om}{\omega}

\newcommand{\Del}{\de}

\newcommand{\GeV}{\,\mbox{GeV}}
\newcommand{\MeV}{\,\mbox{MeV}}

\newcommand{\MSbar}{\overline{\text{MS}}}

\newcommand{\Wcov}{\De}

\newcommand{\beV}{{b_{V}}}
\newcommand{\beRS}{{b_{\RS}}}
\newcommand{\RS}{\vartheta}

\newcommand{\EQ}{Eq.~}

\newcommand{\TAB}{Tab.~}

\newcommand{\FIG}{Fig~}
\newcommand{\FIGs}{Figs~}
\newcommand{\SEC}{section~}

\newcommand{\APP}{appendix~}
\newcommand{\APPs}{appendices~}
\newcommand{\REF}{Ref.~}

\newcommand{\Lag}{{\cal L}}

\newcommand{\dphi}{d_\varphi}

\newcommand{\gone}{A}
\newcommand{\gtwo}{ D}

\newcommand{\gJ}{J}
\newcommand{\ferm}{N}

\newcommand{\Dla} {\overset{\leftarrow}{D}}
\newcommand{\Dra} {\overset{\rightarrow}{D}}

\newcommand{\Lageff}{\Lag_{\text{eff}} }

\newcommand{\XX}{\chi}

\newcommand{\gast}{\ga_*}

\newcommand{\ORD}{{\cal O}}

\newcommand{\scal}{\phi}

\newcommand{\Tud}[2]{T^{#1}_{\;\; #2}}
\newcommand{\TEMT}{ \Tud{\rho}{\rho} }

\newcommand{\mink}{\eta}

\newcommand{\Op}{{\cal O}}

\newcommand{\rsig}[1]{r_{#1}}

\newcommand{\rsg}{r_{\sig,g}}

\definecolor{violet}{rgb}{0.94, 0.2, 0.8}
\definecolor{lightblue}{rgb}{0.39, 0.58, 0.93} 
\definecolor{lightgreen}{rgb}{0.1, 0.73, 0.33}

\DeclareOldFontCommand{\tt}{\normalfont\ttfamily}{\mathtt}
\DeclareOldFontCommand{\bf}{\normalfont\bfseries}{\mathbf}
\DeclareOldFontCommand{\rm}{\normalfont\rmfamily}{\mathrm}

\begin{document}

\title{\boldmath  Gluon Gravitational $ D$-Form Factor: The $\sigma$-Meson \\
 as a Dilaton
Confronted  with  Lattice Data II   
   }

\author[1,2]{Roy Stegeman,}
\author[1]{Roman Zwicky,}

\affiliation[1]{Higgs Centre for Theoretical Physics, School of Physics and
Astronomy, The University of Edinburgh, 
Peter Guthrie Tait Road, Edinburgh EH9 3FD, Scotland, UK}
\affiliation[2]{Quantum Research Centre, Technology Innovation Institute, Abu Dhabi, UAE}
\emailAdd{r.stegeman@ed.ac.uk}
\emailAdd{roman.zwicky@ed.ac.uk}

\abstract{
We investigate the gluon gravitational form factors of the $\pi$, $N$, $\rho$, and $\Delta$
using lattice QCD data at $m_\pi \approx 450 \text{MeV}$ and $m_\pi \approx 170 \text{MeV}$.
We base the analysis on fits to a simple $\sigma/f_0(500)$-meson pole, supplemented by a polynomial background term.
The fitted residues agree with predictions from dilaton effective theory, in which the $\sigma$-meson
acts as the dilaton, the pseudo Goldstone boson of spontaneously broken scale symmetry.
We derive new dilaton-based predictions for the $\rho$- and $\Delta$-gravitational form factors,
and comment on the $\eta_{c}$- and $\eta_b$-form factors in the context of the dilaton interpretation.
These results reinforce our earlier findings,  based on lattice total (quark and gluon) gravitational form factors,
and provide further evidence that QCD dynamics may be governed by an infrared fixed point.}

\maketitle

\setcounter{page}{1}

\mdseries

\section{Introduction}

Gravitational form factors encode the distribution of mass and momentum inside hadrons and have 
become an intensely studied topic \cite{Polyakov:2018zvc,Burkert:2023wzr,Lorce:2024ipy,Lorce:2025oot} in nuclear particle physics.
Driven by new experimental accessibility \cite{Burkert:2018bqq,Duran:2022xag}, recent lattice QCD studies 
\cite{Shanahan:2018pib,Pefkou:2021fni,Hackett:2023rif,Hackett:2023nkr,Wang:2024lrm,Abbott:2025irb} now provide first-principles access to the quark and gluon contributions to the energy-momentum tensor 
for a variety of hadrons.  
\com{Other approaches include 
chiral perturbation theory at low $q^2$ \cite{Belitsky:2002jp,Epelbaum:2021ahi,Alharazin:2020yjv,Alharazin:2023zzc} and light-cone sum rules
at higher $q^2$ \cite{Anikin:2019kwi,Aliev:2020aih,Tong:2021ctu,Tong:2022zax,Ozdem:2022zig,Dehghan:2023ytx,Dehghan:2025ncw,Dehghan:2025eov},
dispersive techniques relying on hadronic input \cite{Broniowski:2024oyk,Cao:2024zlf,Broniowski:2025ctl,Cao:2025dkv},
Dyson-Schwinger methods \cite{Xu:2023izo,Yao:2024ixu,Sultan:2024hep,Xing:2025uwn},  
light-front methods \cite{Xu:2024hfx,Dwibedi:2026ozl}, 
Skyrme-based models \cite{Cebulla:2007ei,Jung:2013bya,Kim:2020lrs,Tanaka:2025pny,Fukushima:2025jah}, 
light-front quark models \cite{Chakrabarti:2020kdc,Choi:2025rto},
and holographic models \cite{Mamo:2021krl,Mamo:2022eui,Fujita:2022jus,Sugimoto:2025btn,Tanaka:2025znc}.}

Besides mapping the hadronic structure through these form factors, it is equally important to understand the dynamical mechanisms that generate the masses they encode. This motivates a complementary line of investigation:
the proposal that strong interactions may be governed by an infrared fixed point \cite{Isham:1970gz,Zumino:1970tu,Ellis:1971sa}, recently reexamined in QCD, both in low-energy processes  
\cite{Crewther:2013vea,Crewther:2015dpa} and more formally \cite{Zwicky:2023bzk,Zwicky:2023krx,Zwicky:2025moo}, by matching scaling dimensions of the underlying theory to those of the effective theory.
Consistency regarding the quark-mass anomalous dimension with lattice simulations 
\cite{DelDebbio:2015byq,Hasenfratz:2020ess,Fodor:2017gtj,LatticeStrongDynamics:2018hun,LatticeStrongDynamics:2023bqp,Bennett:2024tex,LatKMI:2025kti}
(or fits thereto \cite{Appelquist:2017wcg,Appelquist:2017vyy}),
phenomenological models, lower dimensional models \cite{Cresswell-Hogg:2025kvr} and ${\cal N} =1$
supersymmetric gauge theories \cite{Zwicky:2023krx,Shifman:2023jqn} has been established.  
In the wider picture a $\sig$-meson, turning into a dilaton for massless quarks, could provide inspiration for
non-perturbative approaches to QCD in general, be used as a Higgs candidate 
 \cite{Matsuzaki:2012xx,Dietrich:2005jn,Cata:2018wzl,Zwicky:2023krx} or explain various phenomena in nuclear physics \cite{Brown:1991kk,Rho:2021zwm,Rho:2024jdo}.

The underlying idea of the scenario is that spontaneous scale symmetry breaking generates hadron masses.
The dilaton, the Goldstone boson of spontaneous scale symmetry breaking, then couples to hadron masses in such a way that the conformal Ward identities 
are satisfied.\footnote{Examples of exact massless dilatons and Goldstone bosons are rare. Known examples typically arise either in lower-dimensional settings, such as the Gross--Neveu--Yukawa theory~\cite{Cresswell-Hogg:2025kvr}, or in holographic models, where they have been associated with either large condensates~\cite{Elander:2017cle,Elander:2017hyr} or proximity to weak (second-order) phase transitions~\cite{Elander:2020ial,Faedo:2024zib,Elander:2025fpk}.}

In our previous work we found support for  the dilaton interpretation \cite{Stegeman:2025sca},  for the (total) gravitational 
form factors based on $m_\pi \approx 170\MeV$  lattice data~\cite{Hackett:2023rif,Hackett:2023nkr}. 
We return to an earlier lattice study of the \emph{gluon} gravitational form factors at $m_\pi \approx 450\MeV$~\cite{Pefkou:2021fni},
and also make use of the corresponding $m_\pi \approx 170\MeV$ data.\footnote{Although the gluon- and quark-parts are not physical by themselves, they are of phenomenological interest since in 
scattering experiments one contribution typically dominates over the other.}  
Consequently, for a meaningful comparison we determine the gluon-fraction in the same scheme and scale as the lattice study. 
Moreover,   simulations at higher pion mass are computationally less demanding while still providing quantitatively meaningful physical insight. The fact that the $\sigma$,  $\rho$ and  $\De$ become stable is an additional benefit.
This  permits an exploration of the dilaton hypothesis across hadrons of  
spin-$0$ to spin-$\tfrac{3}{2}$  
via the $\pi$, $N$, $\rho$ and $\De$.
The mechanism is tested through the $D$-form factor, as it couples 
to the $\sig/f_{0}(500)$-meson, the dilaton candidate in QCD. 
Explicit scale symmetry breaking due to the quark masses, relevant to  the lattice study,  is assessed at leading order (LO).

The relevant form factors are defined by\footnote{\label{foot:conv} For the $\rho$ and the $\De$ we have only indicated the structures important for this paper.
We refer to \cite{Polyakov:2019lbq,Pefkou:2021fni,Cotogno:2019vjb} and \cite{Kim:2020lrs,Pefkou:2021fni,Cotogno:2019vjb} for a complete decomposition, 
of the six  $\rho$ (seven $\De$) form factors (assuming the symmetric and conserved subset). Note that 
in those references: $A^\rho(q^2)\equiv A_0^\rho(q^2)$, $D^\rho(q^2)\equiv -D_0^\rho(q^2)$, $A^\Delta(q^2)\equiv F_{10}^\Delta(q^2)$ and $D^\Delta(q^2) \equiv F_{20}^\Delta(q^2)$.}
\begin{alignat}{2}
\label{eq:GFF} 
&   \matel{\pi(p')}{T_{\mu\nu}}{\pi(p)}   &\; =  \;&
 a_{\mu \nu} \gone^\pi (q^2) +
d_{\mu\nu} {\gtwo}^\pi (q^2)  \;,  \\[0.1cm]
&   \matel{\ferm(p',s')}{T_{\mu\nu}}{\ferm(p,s)} &\; =  \;&
  \frac{1}{2 m_\ferm}  \bar u(p',s') ( a_{\mu\nu}  \gone^\ferm (q^2)  +
   j_{\mu\nu}  \,  \gJ^\ferm(q^2)  +
  d_{\mu\nu} {\gtwo}^\ferm (q^2) ) u(p,s) \;, \nonumber  \\[0.1cm]
& \matel{\rho(p',\la')}{T_{\mu\nu}}{\rho(p,\la)}  &\; =  \;&   \eps_{\al'}^*(p',\la')\eps_\al(p,\la)(
 - \mink^{\al\al'} (  a_{\mu \nu} \gone^\rho (q^2) +
d_{\mu\nu} {\gtwo}^\rho (q^2)  )  + \dots)   \;,  \nonumber  \\[0.1cm]
 &   \matel{\De(p',s')}{T_{\mu\nu}}{\De(p,s)} &\; =  \;&
  \frac{1}{2 m_\De}  \bar u_\al(p',s') (
 - \mink^{\al\al'} (  a_{\mu \nu} \gone^\De (q^2) +
d_{\mu\nu} {\gtwo}^\De (q^2))  + \dots)   u_{\al'}(p,s)   \nonumber  \;,
\end{alignat}
and depend on the momentum transfer $q \equiv p'-p$ and the momentum average 
${\cal P} \equiv \frac{1}{2}(p + p')$, 
with Lorentz structures ($\sig_{\mu \nu} \equiv \frac{i}{2} [ \ga_\mu,\ga_\nu]$ and $\sig_{\nu q} = \sig_{\nu \mu}q^\mu$)
\begin{equation}
a_{\mu\nu} =2 {\cal P}_\mu {\cal P}_\nu \;, \qquad d_{\mu\nu}  =\frac{1}{2} (q_\mu q_\nu-  q^2 \eta _{\mu\nu} ) \;,
 \qquad j_{\mu\nu} =  i \, {\cal P}_\mu \sig_{ \nu q } + \{\mu \leftrightarrow \nu \} \;,
\end{equation}
that ensures translational invariance, $\partial_\mu T^{\mu\nu} = 0$. 
Spin polarisations are encoded in the Dirac spinor $u(p,s)$, the polarisation vector $\eps_\al(p,\la)$, 
and the Rarita-Schwinger spinor $u_\al(p,s)$, respectively.  
An extensive discussion for generic spin can be found in \REF\cite{Cotogno:2019vjb}.

\begin{figure}[t]
  \centering
    \includegraphics[width=0.7\linewidth]{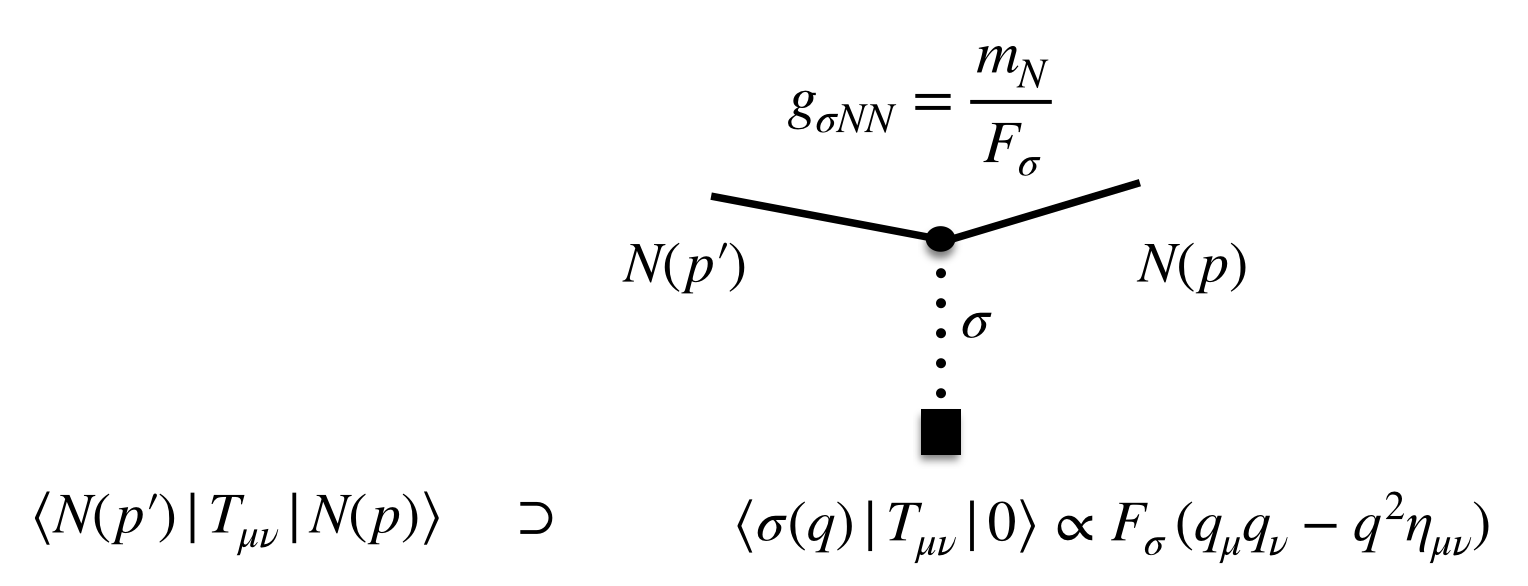}
  \caption{\small Illustration of the dilaton Goldberger-Treiman mechanism.}
  \label{fig:GFF}
\end{figure}

For the subsequent discussion, we define a meaningful nucleon trace 
$\Theta^{ N}  \equiv  \Theta^{N\rho }_{\rho}$ through 
\begin{equation}
\frac{1}{2m_N} \bar u(p',s) \Theta_{\mu\nu}^N(q)  u(p,s)  = \matel{\ferm(p',s)}{T_{\mu\nu}}{\ferm(p,s)} \;.
\end{equation}
The normalisation $A(0)=1$ holds model-independently for all hadrons 
since the (symmetric) energy-momentum tensor $T_{\mu\nu}$ is the Noether current of translation, $P_\mu = \int d^3 x T_{\mu 0}$.
If all form factors remain finite in the soft limit, this implies the textbook formula
$ \Theta^N(0)= 2 m_N^2$ \cite{Donoghue:1992dd},  
and likewise for all other particles.

In a conformal field theory ``$\TEMT=0$'' holds for any physical states and momentum transfer, and 
the same applies when this conformal symmetry is spontaneously broken despite the presence 
of stable massive one-particle states. 
It is the dilaton that realises the conformal Ward identity in
analogy to the role of the pion in restoring the chiral Ward identity in the Goldberger-Treiman  mechanism \cite{Goldberger:1958tr}.\footnote{The effect is also present in the axial-singlet channel 
\cite{Shore:1990zu,Tarasov:2020cwl,Tarasov:2021yll,Tarasov:2025mvn} where though 
the $\eta'$ acquires  a mass from the chiral anomaly. See  \cite{Bhattacharya:2022xxw,Bhattacharya:2023wvy} for a related 
$0^{++}$-analysis where a similar fate is speculated on.} 
For the nucleon, as illustrated in \FIG\ref{fig:GFF} and explained in more detail \SEC\ref{sec:FF}, it requires  
a specific residue of the dilaton pole in 
\begin{equation}
\label{eq:DN}
D^N(q^2) = \frac{4}{3} \frac{m_N^2}{q^2} +   \ORD(1)  \;,
\end{equation}
as proposed  in \cite{Gell-Mann:1969rik}, verified in~\cite{DelDebbio:2021xwu} using dispersion theory 
and  the effective theory in~\cite{Zwicky:2023fay}.

In QCD, which exhibits renormalisation group flow, we do not expect conformality at all scales. 
It is, however, logically possible that as the  quark masses are sent to zero that 
 the trace of the energy-momentum tensor 
\begin{equation}
\label{eq:Theta}
 \Theta^N(0)|_{m_{q} \to 0}  \;,
\end{equation}
vanishes in the soft limit.

The goal of this paper, as in our previous study \cite{Stegeman:2025sca}, is to find remnants of
this mechanism in the lattice data for finite quark masses.
The investigations remain of interest even if the $\sig$ is a pseudo dilaton; massive in the chiral $m_{uds} \to 0$ limit 
but coupling similarly to the massless dilaton.  
Whereas we cannot directly test \eqref{eq:Theta} since finite-quark-mass lattice simulation  induce a  $\sig$-mass,
 we can test its  mass-deformed mechanism \eqref{eq:DN} 
\cite{Zwicky:2023fay}
\begin{equation}
\label{eq:DN2}
D^N(q^2) = \frac{4}{3} \frac{\bar{m}_N^2(1 + \ORD(m_q))}{q^2 - m_\sig^2} + \ORD(1) \;,
\end{equation}
by estimating the impact of the quark mass    on the residue and the $\sig$-mass. (The quantity $\bar{m}_N$ is the nucleon mass in the chiral limit.) The residue-corrections   are non-trivial, as they involve anomalous dimensions, as 
the derivations in \SEC\ref{sec:new} will illustrate.

The paper is organised as follows. 
In \SEC\ref{sec:DEFT}, we derive the $D$-form factors in the dilaton effective theory, including soft $m_q$-corrections
and its gluon-fraction.
The numerical predictions of the effective theory are given in \SEC\ref{sec:dilaton}.
In \SEC\ref{sec:fits}, we discuss the fit ansatz and the fit results, comparing them to the dilaton prediction.  
We comment on the qualitative behaviour of heavy-hadron form factors in \SEC\ref{sec:eta}. 
The paper ends with  summary and conclusions in \SEC\ref{sec:conc}.
The $m_\sig$-dependence of the fits  and the gluon-fraction  of the  $m_\pi \approx 170\MeV$  form factors  are given in \APP\ref{app:additional}.

\section{Gravitational Form Factors in  Dilaton Effective Theory}
\label{sec:DEFT}

This section provides the theoretical foundation for interpreting the gluon form factors. We derive the couplings of the 
$\sig$-meson to spin-$1$ and spin-$\tfrac{3}{2}$ states (\SEC\ref{sec:new}), present the explicit gravitational form factors for all relevant spins (\SEC\ref{sec:FF}), and determine the gluon-fraction from the asymptotics of the renormalisation group equation (\SEC\ref{sec:decay}).
Readers not interested in these technical derivations may skip directly to the final expressions for the form factors in \EQ\eqref{eq:GFF4} and the gluon fraction in \EQ\eqref{eq:xgN}, which are used in the subsequent analysis.

\subsection{Spin-$1$ and spin-$\frac{3}{2}$ dilaton couplings with soft breaking}

\label{sec:new}

Following \REF\cite{Zwicky:2023fay}, we construct the dilaton couplings to vector mesons and spin-3/2 baryons.
 This includes a soft 
perturbation  of the form $\de \Lag = - \la {\cal O}$  (i.e., irrelevant $\De_{\cal O} < d$) for which we have 
 $\la {\cal O} \to  m_q \bar qq $ in mind.
The Lagrangian is parametrised by
\begin{equation}
\label{eq:Leff}
\de \Lag_{\text{eff}} = \frac{1}{2} g_{\sig VV} \sig V_\mu V^\mu
+ g_{\sig\RS\RS} \sig \bar \RS_\mu \RS^\mu \;,
\end{equation}
where $V$ and $\RS$ represent on-shell particles. The fields
 $V_\mu$  and  $\RS_{\mu\al}$  are of the generic vector and Rarita-Schwinger type   \cite{Rarita:1941mf}.
The latter consists of a vector index $\mu$ and a Dirac index $\al$ which we suppress hereafter.

We start with the kinetic terms without a dilaton. For the spin-$1$ and spin-$\frac{3}{2}$ cases the
  Proca Lagrangian and the Rarita-Schwinger Lagrangian read
\begin{equation}
\Lag_{Proca} = - \frac{1}{4} F_{\mu\nu} F^{\mu\nu} 
- \frac{1}{2} m_V^2 V^2 \;, \quad
\Lag_{\text{RS}} = \bar{\RS}_\mu  ( i \ga^{\mu\rho\nu} \partial_\rho - \ga^{\mu\nu} m       ) \RS_\nu \;,
\end{equation}
with $F_{\mu \nu}  =    \partial_{\mu}  V_\nu  - \partial_{\nu}  V_\mu$ and Dirac $\ga$-matrix expressions ($\{\ga_\mu,\ga_\nu\}=   2 \eta_{\mu\nu}$)
\begin{alignat}{2}
& \ga^{\mu\nu}   &\; =  \;& \ga^{[\mu} \ga^{\nu]}=  \frac{1}{2!} [ \ga^\mu,\ga^\nu] = \ga^\mu \ga^\nu - \eta^{\mu\nu}    \;, \nonumber \\[0.1cm]
&  \ga^{\mu \rho \nu}    &\; =  \;& \ga^{[\mu} \ga^{\rho}  \ga^{\nu]} = \ga^\mu \ga^\rho \ga^\nu +
\ga^\rho \eta^{\mu\nu} - \ga^\mu \eta^{\rho \nu} - \ga^\nu \eta^{\mu \rho} \;.
\end{alignat}
The Rarita-Schwinger Lagrangian simplifies to
$\Lag_{\text{RS}} \to  \bar{\RS}_\mu  ( i  \slashed{\partial}  -    m       ) \RS^\mu $, 
when using the  on-shell constraints $\partial^\mu \RS_\mu =0$  and  $ \slashed{\RS} =0$
to match  the Lagrangian in \eqref{eq:Leff}. 

\com{The dilaton effective Lagrangian is obtained through three  steps:  replacing partial derivatives with Weyl-covariant derivatives $\partial_\mu \to \De_\mu$, introducing the dilaton coset field $\hat{\chi} = e^{-\hat{\sig}}$ ($\hat{\sig} \equiv \sig/F_\sig$) \cite{Isham:1970gz,Zumino:1970tu,Ellis:1971sa}
 and adding soft perturbations by 
  following the matching-procedure in  \cite{Zwicky:2023fay}. } 
 The invariant Lagrangian for the vector and the Rarita-Schwinger field read 
 \begin{equation}
\label{eq:LV}
\Lag_{V} = - \frac{1}{4} \hat{\XX}^{\beV} \De_{[\mu} V_{\rho]} \De^{[\mu} V^{\rho]}  -
\frac{1}{2} \hat{\XX}^{\beV} ( \hat{\XX}^{2} \bar{m}_V^2 + \hat{\XX}^{\ga_{{\cal O}}} \Del m_V^2)V^2
 \;,
\end{equation}
and
\begin{equation}
\label{eq:LRS}
\Lag_{\RS} =
\bar{\RS}_\mu  \hat{\XX}^{\beRS} ( i \slashed{\Wcov}
-  \hat{\XX}^{2} \bar{m}_\RS -  \hat{\XX}^{\ga_{{\cal O}}} \Del m_\RS)  \RS^\mu  \;,
\end{equation}
respectively, 
where $\beV = 2(d_V - \cw_V)$ and $\beRS = 2(d_\RS - \cw_\RS)$ are convenient
shorthands,  $\cw_{V,\RS}$   generic Weyl-weights and $d_{V,\RS}$ the engineering dimension of the fields.
The soft-terms $\Del m$ are defined as 
\begin{equation}
m^2_{V} = \bar{m}^2_{V} +  \Del m^2_{V} \;, \qquad
 m_{\RS} = \bar{m}_{\RS} +  \Del m_{\RS}  \;,
\end{equation}
with barred quantities corresponding to the $\la \to 0$ limit, and to linear order in $\la$ one has
\begin{equation}
\Del m^2_{V}  =  \la  \matel{V}{{\cal O}}{V}    \;,\qquad
\Del m_{\RS}  =  \frac{\la}{2 m_\RS}  \matel{\RS}{{\cal O}}{\RS}   \;,
\end{equation}
the same expression as in  \cite{Zwicky:2023fay} for both the boson and the fermion case. 
 It remains to explain the Weyl-covariant derivative, apply it to the fields and then deduce 
 the trilinear couplings $g_{\sig HH}$.

The Weyl-covariant derivative, for a field with Weyl-weight $\omega_\phi$ and 
corresponding Lorentz generator $\Sig_{\mu \nu}$,
 reads  \cite{Mack:1969rr,Isham:1970gz,Isham:1970xz}
\begin{equation}
\label{eq:Delta}
\Delta_\mu \scal  = \partial_\mu + (\cw_\scal \eta_{\mu\nu} + i \Sig_{\mu\nu})  (\partial^\nu \hat{\sig}  )  \scal \;.
\end{equation}
In practice this means that under a Weyl transformation  
$g_{\mu\nu} \to e^{-2 \al} g_{\mu\nu}$,  $\hat{\chi} \to e^{\al} \hat{\chi}$    ($\hat{\sig} \to \hat{\sig} -  \al ) $,
the covariant derivative acting on $\scal$ transforms homogeneously 
\begin{equation}
\Delta_\mu \scal  \to e^{\al(1 + \om_\scal)} \Delta_\mu \scal \;.
\end{equation}
 We note that the dilaton 
 assumes the role of the Weyl-gauge field:
$\partial \hat{\sig} \to \partial \hat{\sig}- \partial \al$. 

In order to apply the Weyl-covariant derivative 
we need the  spin-$\tfrac{1}{2}$ and spin-$1$ Lorentz generators
\begin{equation}
 \Sig_{\mu \nu} \ferm   = \frac{i}{4}[\ga_\mu,\ga_\nu] \ferm \;, \qquad
  (\Sig_{\mu\nu}  V)_\rho =  i \eta_{\mu \rho} V_\nu - \{\mu\leftrightarrow \nu\} \;,
\end{equation}
from which the spin-$\frac{3}{2}$ follows by addition.  Explicitly, we obtain
\begin{alignat}{2}
\label{eq:Weyl132}
& \De_{[\mu} V_{\rho]} &\; =  \;& \partial_{[\mu} V_{\rho]} + (\cw_V-1) \partial_{[\mu}\hat{\sig} V_{\rho]}
 \;, \nonumber \\[0.1cm]
& \slashed{\De} \RS_\rho &\; =  \;& \slashed{\partial} \RS_\rho +
 \cw_{\RS} \slashed{\partial} \hat{\sig} \RS_\rho - \frac{3}{2} \ga_\rho \partial \hat{\sig} \cdot \RS  +
( \partial_\rho \hat{\sig} + \frac{1}{2} \ga_\rho \hat{\sig} ) \slashed{\RS}   \;.
\end{alignat}
For vector mesons, the covariant and the  partial derivative only differs  when 
$\cw_V \neq 1$. Once all on-shell constraints  are applied,
the spin-$\tfrac{3}{2}$ Weyl-derivative reduces to the partial derivative
$ \bar{\RS}^\rho  \slashed{\De} \RS_\rho \to
\bar{\RS}^\rho  \slashed{\partial} \RS_\rho $.
 For the second term in \eqref{eq:Weyl132},  one must use
$\partial_\mu ( \bar  \RS_\rho  \ga^\mu  \RS^\rho )= 0 $, which holds because this quantity is the conserved baryon number current.

Finally, we are in a position to obtain the  on-shell couplings \eqref{eq:Leff}.
For the vector meson it reads
\begin{alignat}{2}
& g_{\sig VV}(q^2)  &\;=\;&  \frac{1}{F_\sig} ( \big[ - \beV (m_V^2 -  \frac{1}{2}q^2) + q^2 ( \cw_{V}-1) \big]
+ (\beV+2) \bar{m}_{V}^2 +  (\beV +  \ga_{\cal O}) \Del m_{V}^2 )   \nonumber \\[0.1cm]
& &\;=\;&
\frac{1}{F_\sig} (2  \bar{m}_{V}^2 +  \ga_{\cal O}  \Del m_{V}^2  + (\dphi-1) q^2 )   \;,
\end{alignat}
where $d_V = \dphi$ was assumed which holds in any dimension. 
 The spin-$\frac{3}{2}$ coupling   reads
\begin{equation}
  g_{\sig\RS\RS} =  \frac{1}{F_\sig} (  - \beRS m_\RS
+ (\beRS+1) \bar{m}_{\RS} +  (\beRS +  \ga_{\cal O}) \Del m_{\RS} )    =  \frac{1}{F_\sig} (\bar m_\RS + \ga_{{\cal O}} \Del m_\RS)  \;.
\end{equation}
The final results exhibit notable similarities, particularly in the cancelation of 
 the Weyl-weights  from 
the final expressions. 
In the chiral limit, this must occur; otherwise the dilaton Goldberger-Treiman mechanism
would not be operative. However, for the $\Del m$-correction it is not obvious why it does.

Similarity to the spin-$0$ and spin-$\tfrac{1}{2}$ cases  \cite{Zwicky:2023fay}
is evident. For comparison, we reproduce all cases here 
\begin{alignat}{4}
\label{eq:g3}
&   \text{spin-}0\!:  \quad  & g_{\sig\varphi\varphi} \; =  \;&
 \frac{1}{F_\sig} (2  \bar{m}_{\varphi}^2 +  \ga_{\cal O}  \Del m_{\varphi}^2  + \dphi q^2)  \;, \quad
& &  \text{spin-}\frac{1}{2}\!:  \quad    &g_{\sig\ferm\ferm}   \; =  \;&
\frac{1}{F_\sig} (\bar m_\ferm + \ga_{{\cal O}} \Del m_\ferm) \;, \nonumber \\[0.1cm]
&   \text{spin-}1\!:    &g_{\sig VV} \; =  \;& \frac{1}{F_\sig} (2  \bar{m}_{V}^2 +  \ga_{\cal O}  \Del m_{V}^2
+  (\dphi-1) q^2   )  \;, \quad
& &  \text{spin-}\frac{3}{2}\!:   \quad & g_{\sig\RS\RS} \; =  \;&  \frac{1}{F_\sig} (\bar m_\RS + \ga_{{\cal O}} \Del m_\RS) \;.
\end{alignat}
We observe that the mass term in the chiral limit is universal
 for bosons and fermions, dictated
by the conformal Ward identity, as mentioned above. The linear-perturbation is also the same
in all cases. Corrections in $q^2$ can only appear for bosons due to their kinetic term.

\subsection{Explicit gravitational form factors for all relevant spins}
\label{sec:FF}

Taking   the $\De$-baryon as an example, we show in more detail how the $D$-form factor appears from its
locally Weyl-invariant Lagrangian
\begin{equation}
\label{eq:Leff2}
\Lageff =  \frac{1}{2} F_\sig^2 ( (\partial \hat{ \XX})^2 -  \frac{1}{6} \,  R \, \hat{\XX}^{2}  )
\; +  \bar{\RS}_\mu  \hat{\XX}^{\beRS} ( i \slashed{\Wcov}
-  \hat{\XX} \bar{m}_\RS -  \hat{\XX}^{\ga_{{\cal O}}} \Del m_\RS)  \RS^\mu  \;,
\end{equation}
and how the on-shell coupling  \EQ\eqref{eq:g3} come into play.
The term proportional to the Ricci scalar is known as the improvement term \cite{Zwicky:2023fay},
while irrelevant to flat-space scattering, it contributes to the energy-momentum tensor because it
defines a coupling to gravity
\begin{equation}
\label{eq:TRdef}
T_{\mu\nu} \supset   T^R_{\mu\nu} =   \frac{F_\sig^2}{6} (\mink_{\mu\nu} \partial^2 -  \partial_{\mu} \partial_{\nu} )   \hat{ \XX}^{2} \;.
\end{equation}
Since the  dilatation current reads  $J^D_\mu = x^\nu T_{\mu\nu}$, with divergence 
$\partial \cdot J^D = \TEMT$,  
the partially conserved dilaton current is analogous to the partially conserved axial current.  As such it implements the Goldstone matrix element
\begin{equation}
\label{eq:Gme}
 \matel{0}{  T_{\mu\nu}  }{\sig} =  \matel{0}{  T^R_{\mu\nu}  }{\sig}
=  \frac{F_{\sig }}{3} ( m_{\sig}^2 \mink_{\mu\nu} -   q_\mu q_\nu) \;,
\end{equation}
 which defines $F_\sig$ as the order parameter of  spontaneous scale symmetry breaking. 
 {It is worth emphasising that low-energy conformal symmetry uniquely fixes the improvement term, thereby removing this ambiguity, which is sometimes viewed as an issue in defining the 
$D$-form factor \cite{Maynard:2024wyi}. }

  The crucial point is that  the linear term in $\sig$ generates the characteristic Goldstone pole. Schematically this reads
  \begin{equation}
    \matel{\RS(p') }{ T_{\mu\nu}  }{\RS(p)} \propto  \matel{0}{  T^R_{\mu\nu}  }{\sig} \frac{1}{q^2 - m_\sig^2} 
    \vev{\sig(q)\RS(p') | \RS(p)} \;,
  \end{equation}
  as illustrated in \FIG\ref{fig:GFF} for the nucleon.
Explicitly,   the actual  LO computation gives
 \begin{equation}
   \matel{\RS(p') }{ T^R_{\mu\nu}  }{\RS(p)} =
 \frac{1}{2m_{\RS}} \bar{u}_i(p,s)   d_{\mu\nu}  \frac{4 m_\RS F_\sig}{3}  \frac{ \, g_{\sig\RS\RS}(q^2) }{q^2-m_\sig^2} u_i(p',s') \;,
 \end{equation}
with  $i$ being any  fixed spatial Lorentz index.
The $D$-form factor can be read off by comparison with the definition \eqref{eq:GFF}.

Finally, assuming the actual particle names, $V,\RS \to \rho,\De$, further adding the pion and 
the nucleon case  \cite{Zwicky:2023fay,Stegeman:2025sca}, one gets
\begin{alignat}{4}
\label{eq:GFFd}
& D^\pi(q^2) &\; =  \;&  \frac{2 F_\sig}{d-1}  \frac{  g_{\sig \pi\pi }(q^2)  }{q^2-m_\sig^2} - 1 \;,  \qquad \quad 
& & D^N(q^2)&\; =  \;& \frac{4 m_N F_\sig}{d-1}  \frac{ \, g_{\sig NN}(q^2) }{q^2-m_\sig^2}  \;,
\nonumber  \\[0.1cm]
& D^\rho(q^2) &\; =  \;&  \frac{2 F_\sig}{d-1}  \frac{  g_{\sig \rho\rho }(q^2)  }{q^2-m_\sig^2} - 1 \;,  \qquad \quad 
& & D^\De(q^2)&\; =  \;& \frac{4 m_\De F_\sig}{d-1}  \frac{ \, g_{\sig\De\De}(q^2) }{q^2-m_\sig^2}  \;,
\end{alignat}
where the bosons show the characteristic factor of $-1$ originating from the kinetic term.
In the pion case the trilinear coupling reads $g_{\sig \pi\pi}(q^2)  =  \frac{1}{F_\sig}  \dphi q^2$ \cite{Zwicky:2023fay},
which  differs from the generic scalar  $g_{\sig\varphi\varphi} $ in \eqref{eq:g3} since 
its pion-mass contribution  must cancel  to comply with a soft-pion theorem  \cite{Zwicky:2023fay}. 
Focusing on $d=4$, adapting the fixed-point anomalous dimension $\ga_{\Op} = \ga^*_{\bar qq} =  - \ga_{m_q}^* = -1$ 
 \cite{Zwicky:2023bzk,Zwicky:2023krx,Zwicky:2023fay} 
 and collecting all the results, including the couplings in \eqref{eq:g3}, we find
\begin{alignat}{4}
\label{eq:GFF4}
& D^\pi(q^2) &\; =  \;&    \frac{  r_\sig^\pi q^2  }{q^2-m_\sig^2} - 1 \;,  \qquad \quad 
& & D^N(q^2)&\; =  \;&    \frac{  r_\sig^N}{q^2-m_\sig^2}  \;,
\nonumber  \\[0.1cm]
& D^\rho(q^2) &\; =  \;&     \frac{  r_\sig^\rho    }{q^2-m_\sig^2} - 1 \;,  \qquad \quad 
& & D^\De(q^2)&\; =  \;&    \frac{  r_\sig^\De   }{q^2-m_\sig^2}  \;,
\end{alignat}
with residues 
\begin{equation}
\label{eq:rsig}
r_\sig^\pi = \frac{2}{3} \;,\qquad   r_\sig^N  = \frac{4}{3} \bar{m}_N^2  \;, \qquad 
r_\sig^\rho = \frac{4}{3} ( \bar{m}_\rho^2 -  \frac{1}{2} \Del m_\rho^2) \;, \qquad 
r_\sig^\De  = \frac{4}{3} \bar{m}_N^2 \;.
\end{equation}
They are the  formal predictions of LO dilaton effective theory. 
It is these expression plus a simple background that will provide the basis for the fit ansatz and interpretation of the lattice data.
\com{It would be interesting to extend the form-factor series to higher spins, making use of the recent ghost-free construction developed in \cite{Ochirov:2022nqz}.}

\subsection{The  gluon-fraction of the $\sig$-meson decay constant}
\label{sec:decay}

We must account for the fact that the lattice simulation \cite{Pefkou:2021fni} includes only the gluon contribution
of the energy-momentum tensor (as  mentioned in the introduction). 
This is to the nucleon's energy and momentum decomposition \cite{Ji:1994av,Ji:1996ek} 
(with the latter being extensively studied on the lattice  \cite{Borsanyi:2020bpd,Wang:2021vqy,Liu:2021lke,Alexandrou:2024zvn,Alexandrou:2024ozj}).
For a meaningful quantitative comparison, we must therefore supply an estimate, at least for the $\sig$-residue, which we refer to as $\rsig{\sig,g}$.
Fortunately, this can be done indirectly via the decay constant since 
it is clear from dispersion theory (see for instance \cite{Zwicky:2016lka}) or \FIG\ref{fig:GFF},  that one has for any hadron state
\begin{equation}
\label{eq:xgdef}
\rsig{\sig} \propto F_{\sig} \, g_{\sig HH} \;, \qquad \rsig{\sig,g} \propto F_{\sig,g} \, g_{\sig HH} \;,
\end{equation}
where $F_{\sig,g}$ is the gluon component of the decay constant, defined below.
\com{We emphasise that the formula is model-independent, as the dilaton’s properties are completely encoded in the coupling $g_{\sig HH}$.}
Hence
\begin{equation}
\label{eq:xg}
r^H_{\sig,g} = \xx_g \, r^H_{\sig} \;, \qquad \xx_g(\mu) \equiv \frac{ F_{\sig,g} (\mu)}{ F_\sig } \;,
\end{equation}
with $\xx_g$ being  the hadron-universal ratio of gluon-to-total $\sig$-meson decay constant.

To define this quantity we need to introduce the gluon- and quark-parts of the energy-momentum tensor (e.g., \cite{Hatta:2018sqd})
\begin{equation}
\label{eq:Tgq}
T^g_{\al \be} = G_{\al \mu} G^{\mu}_{\be} - \frac{1}{4} \mink_{\al\be}G^2 \;, \qquad
T^q_{\al \be} = \frac{1}{4} \sum_q \bar q i \stackrel{\leftrightarrow}{D}_{ \{ \al} \ga_{\be \} } q 
 \;,
\end{equation}
where $x_{\{\al\be\}} = x_{\al\be} + x_{\be\al}$ with 
$\stackrel{\leftrightarrow}{D}  = \Dra - \Dla$ where $\Dra =( \overset{\rightarrow}{\partial} + i g A)$
and $\Dla =( \overset{\leftarrow}{\partial} - i g A)$.
Inserting them into \eqref{eq:Gme}, define the quark and  gluon decay constants
\begin{equation}
\matel{0}{[ T^a_{\al\be}(\mu)]
 }{\sig} = \frac{F^a_{\sig }(\mu)}{3} ( m_{\sig}^2 \mink_{\al\be} - q_\mu q_\nu) +
\FC^a(\mu) \, m_{\sig}^2 \mink_{\al\be} \;, \quad
a \in \{ g,q \} \;,
\end{equation}
where square brackets denote operator and coupling renormalisation. 
The gluon- and quark-parts of the energy-momentum tensor are separately scale- and scheme-dependent 
and therefore not conserved, as evidenced by the non-transverse $\FC^a$-term. Their sum 
$T_{\al\be} = [T^g_{\al\be}(\mu)]+  [T^q_{\al\be}(\mu)]$ is physical and  conserved which implies 
the exact identity $\FC^g + \FC^q = 0$. 
For the  lattice extraction, the   constant $\FC^a(\mu)$ plays no role  because its form factor analogue $\bar C_a$  
does not interfere. 
 To illustrate this, we write the pion decomposition
\begin{equation}
\label{eq:GFFpi}
\matel{\pi(p')}{[T^a_{\al\be}(\mu)]}{\pi(p)} =
a_{\al\be} \,  \gone^\pi_a (q^2,\mu) +
d_{\al\be}\,  {\gtwo}^\pi_a (q^2,\mu) + 2 m_\pi^2 \eta_{\al\be} \, \bar C^\pi_a(q^2,\mu) \;.
\end{equation}
Since the $D$-form factor is determined from the $q_\mu q_\nu$-structure, the $\eta_{\al\be}$-term containing $\bar C^\pi_a$ does not enter. 

Correspondingly the quark- and gluon-form factors are scale and scheme-dependent. The results in the lattice study are quoted for the $\MSbar$-scheme at $\mu = 2\GeV$.
Hence, we need an estimate of $\xx_g$ for this case. We may do so by taking inspiration from the $A$-form factors. Their evolution equation reads, see for instance \cite{Belitsky:2005qn,Hatta:2018sqd}, 
\begin{equation} \label{eq:evol} \frac{d}{d\ln \mu} \vec{A} = \frac{\al_s}{\pi} \gamma^{(2)} \vec{A} \;, \qquad \gamma^{(2)} = \begin{pmatrix} -\dfrac{4}{3} C_F & \phantom{-} \dfrac{1}{3} n_f \\[8pt] \phantom{-} \dfrac{4}{3} C_F & -\dfrac{1}{3} n_f 
\end{pmatrix} \;, \end{equation}
where $\vec{A}^T = (A_q(q^2,\mu),A_g(q^2,\mu))$ and $\al_s = g^2(\mu)/(4 \pi)$ as usual.
To simplify the presentation, we focus on $q^2=0$ and write $A_a(\mu) \equiv A_a(0,\mu)$, which
are equal to
\begin{equation}
A_g(\mu) = \vev{x}_g \;, \qquad A_q(\mu) = \vev{x}_q \;,
\end{equation}
the gluon $\vev{x}_g$ and the quark-momentum fractions $\vev{x}_q$ of the parton distribution functions (see for example \FIG~3.4 in \cite{NNPDF:2024djq}). This
further implies that
the anomalous dimensions above are the second Mellin moments of the splitting function $\ga^{(2)}_{ij} = \int_0^1 dx x P_{ij}(x)$.

The solution of the renormalisation group equation, subject to the boundary condition $A_g(\mu) + A_q(\mu)=1$, reads 
\begin{equation}
\label{eq:ARG}
A_a(\mu) =A_a(\infty) + (A_a(\mu_0) - A_a(\infty)) \left( \frac{ \al_s(\mu)}{\al_s(\mu_0)} \right)^{\frac{8 C_F+ 2 n_f}{3\be_0}} \;, \quad
a \in \{ g,q \} \;,
\end{equation}
with asymptotic values
\begin{equation}
A_g(\infty) = \frac{4 C_F}{ 4 C_F + n_f} \;, \qquad A_q(\infty) = \frac{n_f}{ 4 C_F + n_f} \;,
\end{equation}
where $C_F = (N_c^2-1)/2N_c$ for an $SU(N_c)$ gauge group, $n_f$ the number of flavours and  
$\frac{d}{d\ln \mu} \al_s = - \frac{\be_0 }{2 \pi} \al_s^2 + \ORD(\al_s^3)$ with
$\be_0 = \frac{11}{3} C_A - \frac{2}{3} n_f$.

The key point is that the momentum fraction, and thus $A_a(\mu)$, is known to be close to their asymptotic
values at intermediate scales. We will focus on the $n_f=3$ case as appropriate for
the $m_\pi \approx 170 \MeV$ lattice simulation with dynamical up, down and strange quark 
\cite{Hackett:2023rif,Hackett:2023nkr}. At $\mu = 2\GeV$ they find
\begin{equation}
A^\pi_g(2 \GeV) =0.546(18) \;, \qquad A^N_g(2 \GeV) = 0.501(27) \;,
\end{equation}
whose averaged central values $\approx 0.52$ lies within $20\%$ of the asymptotic estimate $A_g^\infty = 0.64$ (with $C_F =4/3$).
This corresponds to a deviation just below $14\%$ at $\mu = 6 \GeV$ which might be considered a preasymptotic scale.
Since for local operators the evolution equations are independent of the external states, we have that
the replacement $A^{\pi,N}_g \to \xx_g$ is formally equivalent in \EQ\eqref{eq:ARG}. Moreover, 
since with $\xx_g + \xx_q =1$  the same overall normalisation is shared, this implies the same asymptotic values,
$ \xx_{g}(\infty) = A_g(\infty)$.
We therefore take
\begin{equation}
\label{eq:xgN}
\xx_g(2\GeV) = 0.64(20) \;,
\end{equation}
as our  value used to estimate the gluon residues in \EQ\eqref{eq:xg}. Whereas a priori $\xx_g$
 is unconstrained and could, for example, be negative, its relation to the momentum fractions and its proximity to the asymptotic regime prevent this from happening.

One may also assess our reasoning more directly through the $D$-form factor itself.
It satisfies the same evolution equation \eqref{eq:evol} and thus the asymptotic values are the same
$\xx_{D_g}(\infty) = A_g(\infty) $, for the normalised quantity $ \xx_{D_g} \equiv  D_g/D$.
This suggests  that we should have $\xx_{D_g}(2 \GeV) = \xx_{D_g}(\infty) \pm 20\%$.
As can be inferred from the plots in \FIG\ref{fig:ratio}, this expectation is reasonably well satisfied for the $m_\pi \approx 170 \MeV$ lattice data. 
The plots indicate that at $\mu=2 \GeV$, $\xx_{D_g}$ is slightly larger than $A_g$, closer to the asymptotic value.
This is compatible  with $\sig$-dominance itself by \EQ\eqref{eq:xgN}, and suggest if anything that the true value of $\xx_g$ might be slightly larger than the central value in \eqref{eq:xgN}.
In summary, we find a consistent picture, indicating that \eqref{eq:xgN} is a trustworthy estimate.

\section{Dilaton Residues of the $\pi$, $N$, $\rho$ and $\Delta$ at  $m_\pi \approx 450 \MeV$ }
\label{sec:dilaton}

We now turn to the predictions of the residues given in \eqref{eq:GFF4}, including an uncertainty analysis.
For this purpose, we need the hadron masses at $m_\pi \approx 450 \MeV$, 
which were determined in a separate nuclear physics study for the same ensemble \cite{Orginos:2015aya}. 
Their quoted uncertainties do not include the continuum limit, since only one lattice spacing was used, but this effect is reported to be small.
To crosscheck this statement on the extrapolation, we use the nucleon mass from  \REF\cite{Lin:2019pia}  as a proxy. Although this calculation was performed on a different ensemble, it employed the same method of Clover-Wilson fermions.
Extrapolating from a similar lattice spacing, we find that the deviation from the continuum is approximately 
0.6\%. 
Although \REF\cite{Lin:2019pia}   provides three lattice spacings for continuum extrapolation, we consider only a single one 
to estimate the discretisation effects  in \cite{Orginos:2015aya}.
To account for the increased uncertainty from this single-point estimate, we conservatively double this to $1.2\%$ and add it in quadrature to the other uncertainties (except for the pion, which was used for tuning).
The resulting masses and their uncertainties are summarised in \TAB\ref{tab:450mass}.
\begin{table}[h]
  \centering
  \small
  \setlength{\tabcolsep}{6pt} 
  \renewcommand{\arraystretch}{1.3} 
  \begin{tabular}{l |  r | r | r || r |  r }
    $H$     & $j$           & $m_{H}|_{m_\pi=450\mathrm{MeV}}$~\cite{Orginos:2015aya} & 
    $\bar m = m_{\textrm{H}}|_{m_{uds} \to 0}$  & $ m^{\textrm{PDG}}_{\textrm{H}}$ &   $\Ga^{\textrm{PDG}}_{\textrm{H}}$        \\ \hline
  $\pi$   & $0$           &  $449.9(4.6)$                         & $0$             & $134.98$ & $  \approx 10^{-5}  $  \\
    $N$     & $\frac{1}{2}$ &  $1226(19)$                           & $826$           & $939.565$ & $ 0 $                 \\
  $\rho$  &  $1$          &  $887.3(14.4)$                        & $675(10)$       & $775.26$ & $ 149.1  $              \\
    $\De$ & $\frac{3}{2}$ &  $1486(24)$                           & $1152(25)$     & $ 1232$ & $ 117 $
  \end{tabular}
  \caption{\small All mass scales are in  units of $\MeV$. 
  The masses 
  $m_{H}|_{m_\pi=450\mathrm{MeV}}$ are determined from the lattice QCD ensemble~\cite{Orginos:2015aya}, and
  $\bar{m}$ are the masses in the chiral limit (see text for details).
  For comparison, the last two columns list the PDG~\cite{ParticleDataGroup:2024cfk} masses and widths (of 
   the proton and otherwise uncharged particles). The pion width is small as it is stable in QCD.
  The $\rho$ and $\De$-values in the
  SU(3) chiral limit $m_{uds} \to 0$ are understood as the ones appearing at LO in the effective theory. }
  \label{tab:450mass}
\end{table}


\subsection{The $\pi$-meson residue}

Since  our prediction \eqref{eq:rsig} is  
 independent of the pion mass $m_\pi^2 = \ORD(m_q)$ to linear order,  
 its central value remains the same $r_\sig^\pi = \tfrac{2}{3}$ 
 as in~\cite{Stegeman:2025sca}.  The unknown radiative correction of relative order 
$\ORD( (m_\pi/F_\pi)^2 \ln m_\pi^2 )$ will be increased compared to~\cite{Stegeman:2025sca}.  
We therefore need to estimate $y \equiv (m_\pi/F_\pi)_{450}/(m_\pi/F_\pi)_{170}$ (where the subscript indicates the pion mass). 
Using N$^2$LO chiral perturbation theory~\cite{Bijnens:2017wba}, with the  largely-unknown low-energy constants $c_i$  
set to zero, gives $y \approx 1.8$. 
%
Alternatively, we can estimate it from the $m_\pi$-dependence of $F_\pi$, known from lattice QCD~\cite{Brandt:2013dua}. 
The plots indicate  $(F_\pi)_{450}/(F_\pi)_{170}\approx 1.3$, which gives $ y \approx2.0$.
Hence, both ways give  similar results, indicating consistency. 
Naively, this would quadruple the corrections; however, since chiral logarithms are not sizeable at $m_\pi=450\MeV$ 
and the uncertainty in \REF~\cite{Stegeman:2025sca} was  rather  generous
\begin{equation}
\label{eq:rpiDil}
r_\sig^\pi|_{\text{dilaton}} = \frac{2}{3} \pm 0.2 \;, \quad  \quad  \rsig{\sig,g}^\pi|_{\text{dilaton}} =  0.43(22) \;,
\end{equation}
provides a credible estimate. 
The gluon-residue   is estimated from  
the gluon-fraction  \eqref{eq:xgN}  by adding uncertainties  in quadrature. 

\subsection{The $N$-baryon residue }

As for the pion, the nucleon residue is quark-mass independent.
Thus, we obtain the same value $r^N_\sig=\tfrac{4}{3}\bar{m}_N^2=0.91\GeV^2$ as in \REF\cite{Stegeman:2025sca}, 
based  on the nucleon mass decomposition\cite{Hoferichter:2025ubp}. 
 The uncertainty analysis differs from the pion as the nucleon and the remaining hadrons are not  Goldstones.
We will follow the same strategy as in \cite{Stegeman:2025sca} and use the $m_q$-dependence of the 
hadronic mass for guidance.   For the baryons one only needs to consider one power into the error analysis 
because  the other power  originates from the definition \eqref{eq:GFF}. 
Since the $\ORD(m_q)$ corrections are known, one needs to focus on the remaining ones.  We estimate 
them to be covered by half of the total $m_q$-dependent correction:  $\frac{1}{2}(m_N|_{450} - \bar{m}_N)= \frac{1}{2}(1.125-0.826)=0.15 \GeV$. 
Including  another $10\%$ due to $q^2$-dependent corrections, and adding uncertainties in quadrature gives   
 our final estimate 
\begin{equation}
\label{eq:rNDil}
r_\sig^N|_{\text{dilaton}} =  0.91(18) \GeV^2 \;, \quad  \quad  \rsig{\sig,g}^N|_{\text{dilaton}} =  0.58(22) \GeV^2 \;.
\end{equation}

\subsection{The $\rho$-meson residue  }

The $\rho$-residue \eqref{eq:rsig} splits into a chiral-limit ($m_{uds}\rightarrow 0 $)  and a correction part due to the anomalous dimension. 
From \FIGs 24 and 26 in \cite{Molina:2020qpw}, we infer $\bar m_\rho = 660 (10) \MeV$,   notably smaller 
than  the light-quark  limit $m_\rho|_{m_{u,d}\to 0 } \approx 740\MeV$ (see their \FIG 26 and  also  \cite{Niehus:2020gmf}). 
From  \TAB\ref{tab:450mass}, we find $m_\rho|_{450}=887.3(14.4) \MeV$, leading to a sizeable correction
 $\Del  m_\rho^2 = m_\rho^2|_{450} - \bar m_\rho^2 = 0.35\GeV^2$  and  a central 
value of $r^\rho_\sig \approx 0.35 \GeV^2$.  The uncertainty is estimated as for the nucleon $\tfrac{1}{2}(m_\rho|_{450} - \bar m_\rho) \approx 100\MeV$. Unlike for the fermion, $m_\rho^2$ is entirely dynamical in  \eqref{eq:rsig}. 
Hence,  we take 
$\frac{4}{3}  ( ( \bar{m}_\rho^2 - \frac{1}{2} \Del m_\rho^2 )^{1/2} + 100\MeV)^2 \approx 0.19\GeV^2$ as an  estimate 
of the uncertainty.
Adding again a 10\% error due to the $q^2$-dependence and combining uncertainties in quadrature gives our final estimate
\begin{equation}
\label{eq:rrhoDil}
r_\sig^\rho|_{\text{dilaton}} =  0.35(19) \GeV^2 \;,  \quad  \quad \rsig{\sig,g}^\rho|_{\text{dilaton}} =  0.22(19) \GeV^2 \;.
\end{equation}
The relative  error is rather large due to the significant reduction of the central value by the linear quark-mass term. 
For example, if $\gast = (0,-1)$  then  $ r_\sig^\rho|_{\text{dilaton}}$ would assumes values of $(0.58,0.81)$, respectively.

\subsection{The $\De$-baryon residue  }

The $\De$-residue  \eqref{eq:rsig} is  similar  to the nucleon case.  
The chiral limit  $\bar{m}_\De = 1152(25) \MeV$ can be inferred  from an N$^3$LO chiral perturbation theory calculations~\cite{Ren:2013oaa},  consistent with the $N$-$\De$ mass splitting 
$\bar{m}_\De - \bar{m}_N = 330\MeV$~\cite{Bernard:2005fy} and $317\MeV$~\cite{Pascalutsa:2005nd}.
This leads to a central value of $r^\De_\sig=1.77(8) \GeV^2$.  
As for the nucleon  we estimate the remaining (radiative) corrections to be
$\tfrac{1}{2}(m_\De |_{450} - \bar{m}_\De )=\tfrac{1}{2} (1.486-1.152)\GeV=0.17\GeV$. 
Adding another $10\%$ $q^2$-dependent corrections and adding uncertainties in quadrature we obtain
our final estimate
\begin{equation}
  r^\De_\sig |_{\text{dilaton}}  = 1.77(28) \GeV^2 \;, \quad  \quad  r_{\sig,g}^\De|_{\text{dilaton}} =  1.13(39) \GeV^2 \;.
\end{equation}

\section{Fits to the Lattice Data}
\label{sec:fits}

Having established the theoretical predictions for the dilaton residues, we now compare them to lattice QCD data through systematic fits to  the gluon $D$-form factor   lattice data at $m_\pi \approx 450 \MeV$ \cite{Pefkou:2021fni}  
for the $\pi$, $N$, $\rho$ and $\De$, 
and $m_\pi \approx 170 \MeV$ \cite{Hackett:2023rif,Hackett:2023nkr} for the $\pi$  and $N$.
In the tables and plots the two data sets are distinguished by 
$\pi(450)$ and $\pi(170)$, in self-explicatory manner.

\begin{figure}[t]
  \centering
    \includegraphics[width=0.49\linewidth]{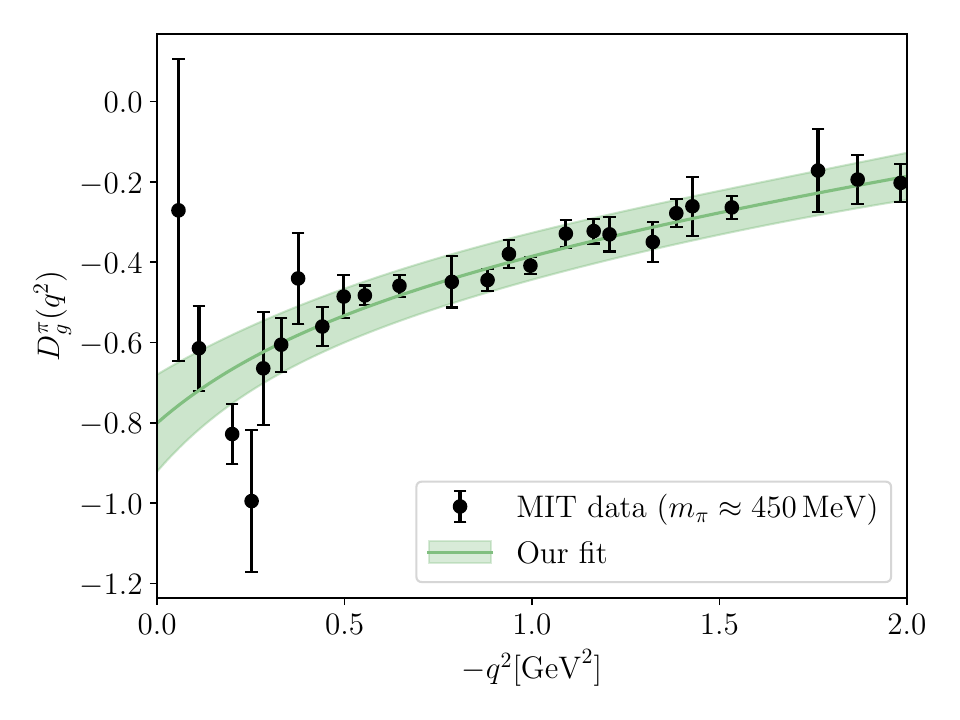}
     \includegraphics[width=0.49\linewidth]{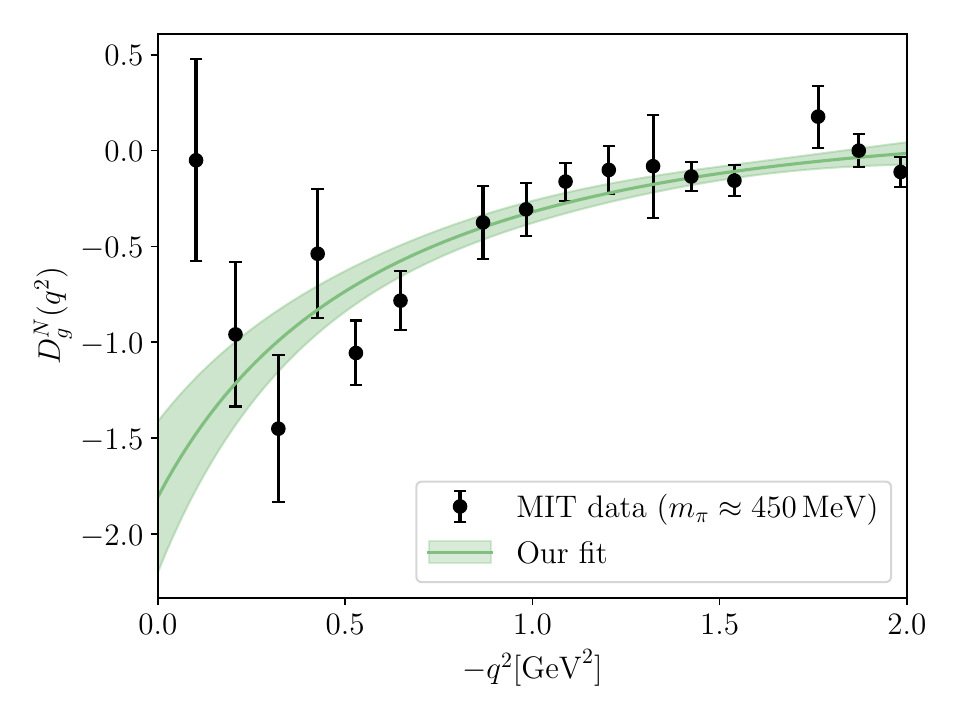}
  \includegraphics[width=0.49\linewidth]{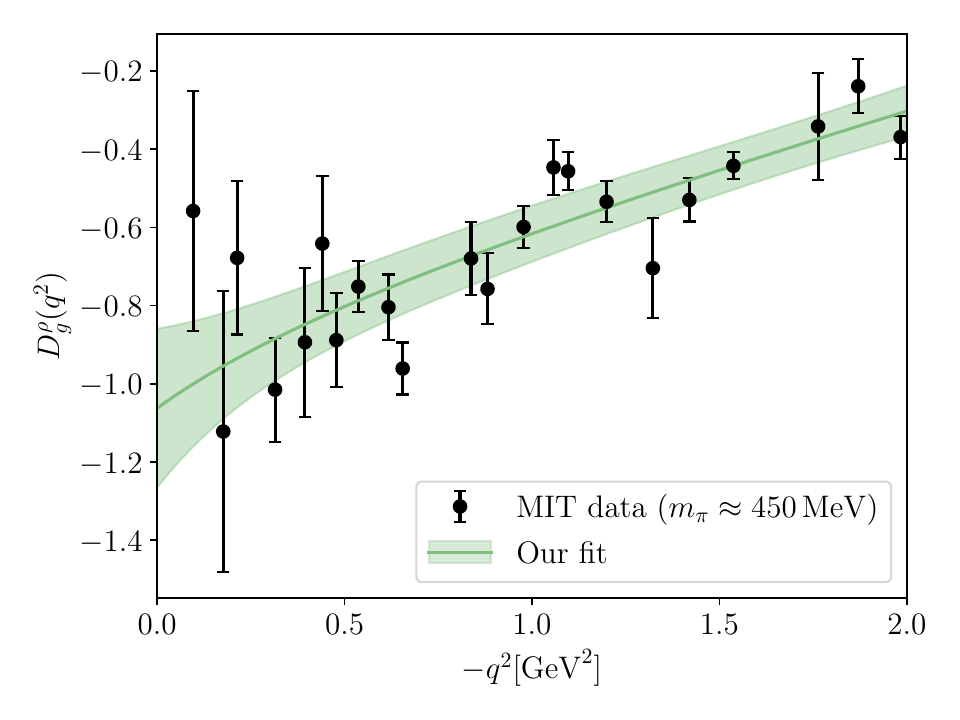}
  \includegraphics[width=0.49\linewidth]{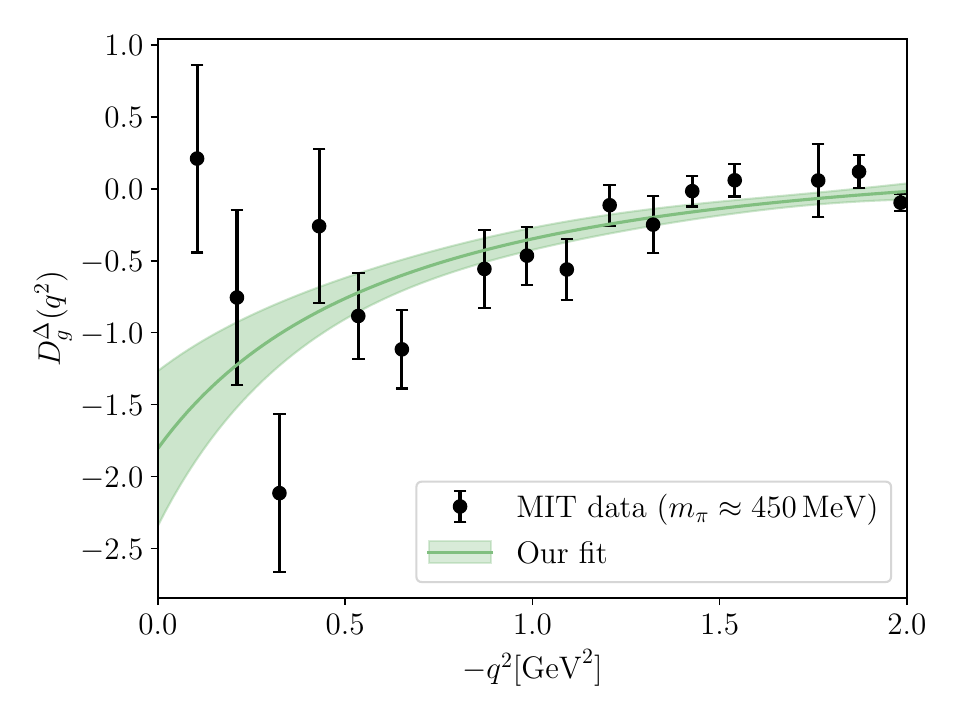}
   \includegraphics[width=0.49\linewidth]{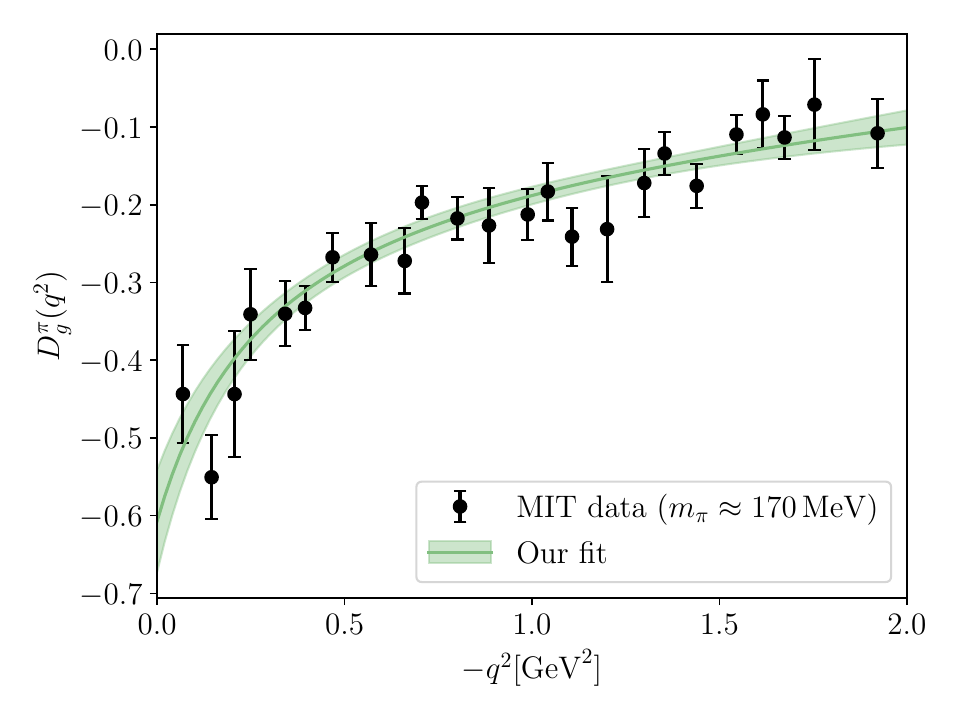}
  \includegraphics[width=0.49\linewidth]{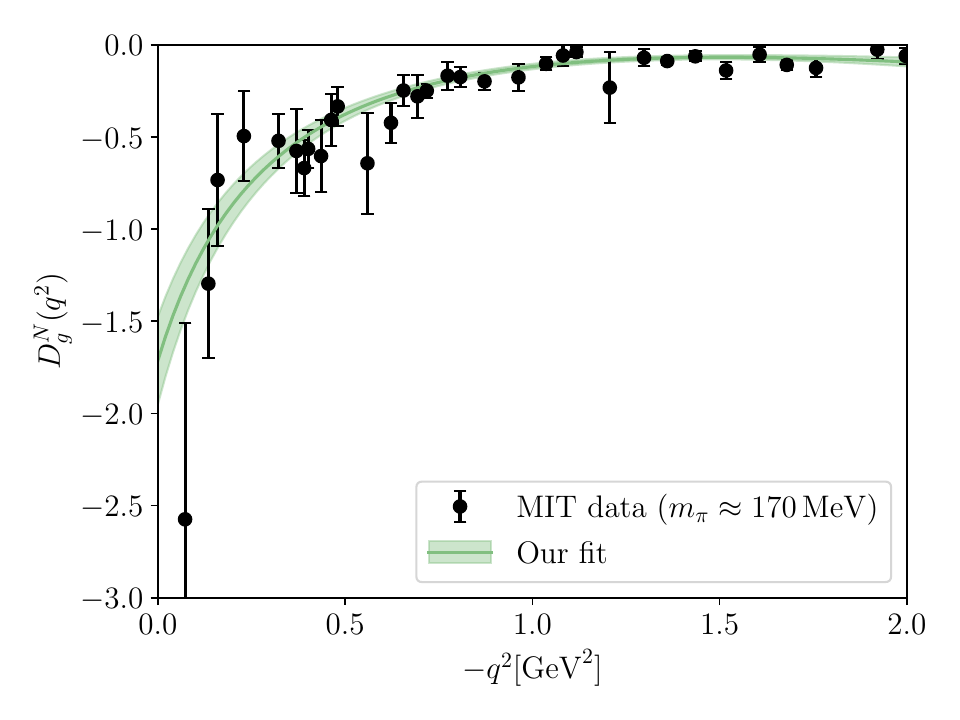}
  \caption{\small Fits using  the ansatz \eqref{eq:Bansatz} for the baryons and \eqref{eq:Mansatz} for the mesons, 
  compared to the MIT lattice QCD data~\cite{Pefkou:2021fni}  (first four) and 
  \cite{Hackett:2023nkr,Hackett:2023rif} (last two).
  Corresponding fit-parameters are given in \TAB\ref{tab:450}. 
  The dark green line is the best-fit result and the light green band denotes the $68\%$-confidence interval. 
   See footnote \ref{foot:conv} for 
  our conventions compared to \cite{Pefkou:2021fni},  explaining the sign-difference in the $\rho$-plot.}
  \label{fig:450_fits}
\end{figure}

We must address what value to 
   the $\sig$-mass takes for $m_\pi \approx 450 \MeV$.  Although it has not been 
 determined in \cite{Pefkou:2021fni}, we can infer it from the lattice studies on  the $\sig$-meson \cite{Briceno:2016mjc,Briceno:2017qmb,Rodas:2023twk}.\footnote{It is a long-term project of the lattice community to determine the physical $\sig$-pole;
 in a series of papers the HadSpec collaboration 
 \cite{Briceno:2016mjc,Briceno:2017qmb,Rodas:2023twk} has reached as low as $m_\pi \approx 239 \MeV$.}
 Specifically, 
 one has $m_\sig|_{m_\pi\approx330{\tiny \MeV}}=657(3)\MeV$ from $m_\sig/m_\pi=1.99(1)$ and 
 $m_\sig|_{m_\pi\approx 391{\tiny \MeV}}=759(8)\MeV$ from $m_\sig/m_\pi=1.94(2)$.
 Performing a linear extrapolation to $m_\pi=450\MeV$ then gives   $m_\sig  = 857(16)\MeV$, not accounting for the uncertainty in the pion masses. 
Inspection of \FIG~\ref{fig:450_fits} shows that the data become less reliable as $q^2 \to 0$, which reflects the fact that it is extracted from the $q_\mu q_\nu$-structure.  To partially account for this effect, we adapt a 
conservative $50\MeV$-uncertainty on the $\sig$-mass
\begin{equation}
\label{eq:850}
m_\sig|_{m_\pi \approx 450{\tiny \MeV}} =850(50)\MeV \;,
\end{equation}
as our estimate of the $\sig$-mass for the $m_\pi \approx 450\MeV$.

For the $m_\pi \approx 170 \MeV$ case we may use the same approach as in  \cite{Stegeman:2025sca}, 
taking into account the gluon fraction $\xx_g$ in \eqref{eq:xgN} which leads to predictions quoted in 
\TAB\ref{tab:450}.  {Despite its unstable nature, the $\sig$-meson is well described in the Euclidean domain by a simple pole with an effective mass $m_\sig = 550(50)\MeV$. In our earlier work \cite{Stegeman:2025sca} (see also the references therein), this feature was analyzed in two complementary ways: (i) through a multipole expansion in momentum space, where the simple pole corresponds to the ``monopole,'' and (ii) via an explicit calculation in the linear $\sig$-model, employed as a \emph{toy model}. Intuitively, this indicates that the detailed resonance structure of the 
$\sig$-meson does not play a significant role in the Euclidean domain, where the form factor remains real-valued. }

 The fit ansatz consists of the LO dilaton predictions \eqref{eq:GFF4} supplemented by a linear polynomial that accounts for higher-state contributions. 
Unlike our previous analysis \cite{Stegeman:2025sca}, we do not attempt to fit a second pole, as the available data are less precise.
For the nucleon, the fit ansatz reads
\begin{equation}
\label{eq:Bansatz}
D^{N}_g(q^2) = \frac{\rsg^{N}}{q^2-m_\sig^2} +b^{N}_g+b^{'N}_g q^2 \;,
\end{equation}
and analogous for the $\De$-baryon.  The main idea is that the background polynomial effectively models higher resonances, as discussed extensively in \REF~\cite{Stegeman:2025sca}. It should not be interpreted as a systematic effective field theory expansion in $q^2/(4 \pi F_\sig)^2$, whose radius of convergence would lie well below the data range. 
Given the quality of the data, we do not attempt to fit alternative background parameterisations. Instead, we take confidence from the fact that the chosen ansatz was tested in \REF~\cite{Stegeman:2025sca} against a variety of background combinations.

The only difference for the mesons \eqref{eq:GFF4} is the constant term of minus one which originates from the 
bosonic kinetic term in the effective theory.  Since it governs the quark and gluon momentum fractions, encoded 
in $A_{q,g}(0)$, it is then clear that the correct amendment to split into a dilaton and background part is
\begin{alignat}{2}
\label{eq:Mansatz}
D^{\pi}_g(q^2)  &= \frac{q^2 \, \rsg^\pi }{q^2 - m_\sig^2}
                  - A_g^\pi(0) + \hat{b}^{\pi}_g + b^{'\pi}_g q^2 \;, \\[4pt] \nonumber
D^{\rho}_g(q^2) &= \frac{\rsg^\rho}{q^2 - m_\sig^2}
                  - A_g^\rho(0) + \hat{b}^{\rho}_g + b^{'\rho}_g q^2 \;.
\end{alignat}
with $A_g^\pi(0) =0.550(47)$ and $A_g^\rho(0) = 0.478(44)$ taken from  the data in \REF\cite{Pefkou:2021fni}. 
This is mainly relevant for the pion, since in that case the soft-theorem dictates $D^{\pi}(0) = -1$  \cite{Donoghue:1991qv,Zwicky:2023fay}
(or  $D_g^{\pi}(0) = -A_g^\pi(0)$ in our case) 
 such that $\hat{b}^\pi_g =0$ ought to hold by consistency.  
 We do not enforce the large $q^2 \to -\infty$ asymptotic form   \cite{Tong:2021ctu,Tong:2022zax} of 
 the $D$-form factors as it is not clear whether this regime is reached 
 \cite{Stegeman:2025sca}.

\subsection{Fit results and interpretation}
\label{sec:fits2}

\begin{figure}[h!]
  \centering
    \includegraphics[width=0.6\linewidth]{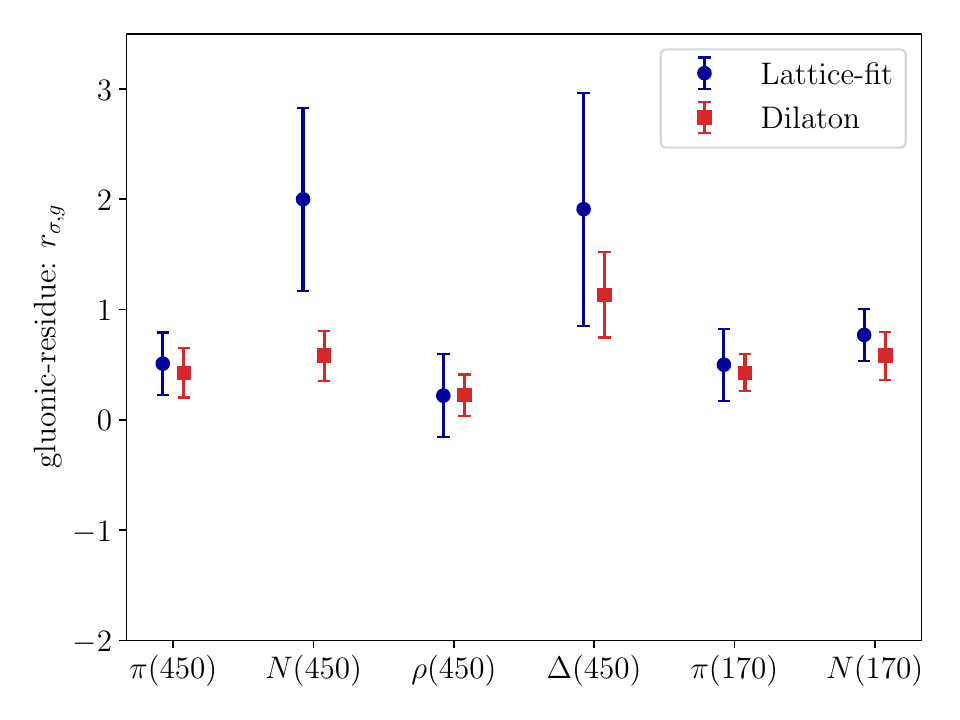}
  \caption{\small  The gluon-residue  fits for the $m_\pi \approx 450 \MeV$ lattice data (first four) and 
  the $m_\pi \approx 170 \MeV$ lattice data (remaining two), compared against the dilaton effective theory predictions, 
  see  \TAB\ref{tab:450}.  To gauge the agreement it is worthwhile to consider the $\chi^2$/d.o.f. in that table.}
  \label{fig:barchart}
\end{figure}

We fit   \eqref{eq:Bansatz} for baryons and 
\eqref{eq:Mansatz} for mesons, with $m_\sig = 850(50)\MeV$ for the $m_\pi \approx 450 \MeV$-data as discussed above
and $m_\sig = 550(50)\MeV$ for the $m_\pi \approx 170 \MeV$ case.
Model uncertainties are taken into account by  $m_\sig$-variation (see \TAB\ref{tab:massvariation2} in the appendix). 
Results are shown in \FIG\ref{fig:450_fits} and 
  \TAB\ref{tab:450}, including the dilaton predictions, and further illustrated in 
\FIG~\ref{fig:barchart} where the fit and $m_\sig$-variation uncertainty are added in  quadrature. 
There is  good agreement between our fits (curves) and the lattice data (points).\footnote{Following \REF~\cite{Pefkou:2021fni} (see in particular \APPs~A.5 and A.6), our parametrisations are fitted to the bare gravitational form factors and subsequently renormalised. This procedure avoids the d'Agostini 
 bias~\cite{DAgostini:1993arp} arising from non-Gaussianity introduced by the renormalisation while a $\chi^2$-fit assumes Gaussian uncertainties. The fit parameters quoted here are the renormalized values.}
  From \TAB\ref{tab:450}, we note that the $\chi^2$ per degree of freedom (d.o.f.)  is close to one and thus excellent  for $\pi(450)$, $\pi(170)$ and $N(170)$ and somewhat higher  $\chi^2$/d.o.f. $\approx 2$ for 
  $N(450)$, $\rho(450)$ and $\De(450)$.  This trend is consistent with the findings of the original works employing di- and tripole fits~\cite{Pefkou:2021fni,Hackett:2023rif,Hackett:2023nkr}, indicating that the $m_\pi \approx 170\MeV$ data is of better quality.   The slight outlier in the $N(450)$ {has a pull of
  $p = (r_{\sig,g}|_{\text{fit}} -  r_{\sig,g}|_{\text{dilaton}})/ (\sig^2_{\text{fit}}+ \sig^2_{\text{dilaton}} )^{1/2}  \approx 1.64$ above 
  the expected normal distribution. This translates to a p-value of $0.1$ and  }   
  is statistically acceptable but also brought into perspective by its 
  larger $\chi^2$/d.o.f. and the excellent agreement of the {combined quark and gluon (i.e. total) $N(170)$-case  \cite{Stegeman:2025sca}.  
  The fit of all results, measured  by the reduced chi-squared
   $\chi^2 = \sum_{i=1}^6 \text{pull}_i^2/6 \approx 0.66$,  corresponds 
  itself to a p-value of $0.68$ indicating consistency.}  
 Inspecting \TAB\ref{tab:450}, one observes that the predictions tend to lie on the lower side, which may be related to the previous observation that the central value of $\xx_g$ in \eqref{eq:xgN} could be slightly underestimated. 
 {This is supported  if we fit the gluon momentum fraction factor 
 $z_g \to y\, z_g $ for which  we obtain $y = 1.34(27)$ with $\chi^2/5 = 2.34/5 \approx 0.47$ 
 with a corresponding p-value of 0.8.}

\begin{table}[t]
  \centering
  \small
  \setlength{\tabcolsep}{6pt} 
  \renewcommand{\arraystretch}{1.2} 
\begin{tabular}{l | rrrr || rr}
  & $\pi(450)$ & $N(450)$ & $\rho(450)$ & $\Delta(450)$ & $\pi(170)$ & $N(170)$ \\ \hline
   $j$ & $0$  & $\tfrac12$ & $1$  & $\tfrac32$ & $0$  & $\tfrac12$ \\ \hline
    \rowcolor{gray!6}  $\rsg|_{\text{dilaton}}$                      & $0.42(24)$   & $0.58(25)$ & $0.22(19)$  & $1.13(39)$  &$0.43(13)$   & $0.58(17)$  \\ \hline
  \rowcolor{gray!20}  $\rsg|_{\text{fit}}$                      & $0.51(28)(3)$   & $2.00(74)(37)$ & $0.22(37)(7)$  & $1.91(98)(39)$  &$0.50(13)(2)$   & $0.77(13)(12)$  \\ \hline 
 \text{pull}  &   $0.24$ & $1.64$ & $0.00$ & $0.69$ & $0.38$ & $0.78$ 
  \\ \hline
  $b_g$ or $\hat{b}_g$        & $-0.25(10)$  & $0.96(66)$ & $-0.27(35)$ & $0.84(86)$  & $-0.067(66)$ & $0.71(17)$  \\
  $b'_g~[\mathrm{GeV}^{-2}]$  & $-0.12(07)$  & $0.12(22)$ & $-0.27(12)$ & $0.08(27)$ & $-0.037(35)$ & $0.230(67)$  \\ \hline
  $\rho_{r,b}$               & $-0.85$      & $0.99$     & $0.98$      & $0.99$  & $-0.94$  & $0.98$   \\
  $\rho_{r,b'}$                & $0.95$       & $0.95$     & $0.95$      & $0.94$    & $0.86$  & $0.92$   \\
  $\rho_{b',b}$               & $-0.74$      & $0.98$     & $0.98$      & $0.98$    & $-0.66$  & $0.98$  \\ \hline
  $N_\mathrm{dat}$            & $25$         & $16$       & $22$        & $16$  & $24 $        & $33$        \\
  $\chi^2/\mathrm{d.o.f.}$    & $1.3$        & $1.8$      & $2.0$       & $2.2$     & $1.06$       & $0.88$    \\
\end{tabular}
  \caption{\small
  The gluon-residue dilaton prediction and fit data
   for the $m_\pi \approx 450 \MeV$  data (first four) and 
  the $m_\pi \approx 170 \MeV$  data (remaining two), for the 
    baryon- and meson-ansatz in 
     \eqref{eq:Bansatz}   and \eqref{eq:Mansatz}, respectively. 
    The first error is the fit uncertainty for fixed $\sig$-mass and the second one is due 
    to the $m_\sig$-variation shown in \TAB\ref{tab:massvariation2}.  
    The pion residues are dimensonless whereas all others  are in units of $\GeV^2$.  Fit-correlations are encoded in $\rho_{x,y}$.}
  \label{tab:450}
\end{table}
 
We turn to the soft-pion prediction, $\hat{b}^\pi_g|_{\text{soft}} = 0$.
The fit result $\hat{b}^\pi_g|_{\text{fit}} = -0.25(10)$  is reasonably compatible  when we take into account that 
the constant value $A_g^\pi(0) = 0.550(47)$ is sizeable and 
relative $20\%$-radiative corrections  due to $m_\pi = 450\MeV$, resulting in $\hat{b}^\pi_g|_{\text{soft}} = 0.0(1)$.
For the $m_\pi \approx 170\MeV$ data, we obtain $\hat{b}^\pi_g = -0.067(66)$, showing excellent agreement. Combined with the equally good $\chi^2$-values for both the 
gluon- and quark-contributions, this again suggests that the $m_\pi \approx 170\MeV$ data are more reliable than the $m_\pi \approx 450\MeV$ data.

In summary,  the fit results, given in \TAB\ref{tab:450}  and displayed in  \FIG\ref{fig:barchart}, 
are consistent with the hypothesis that the $\sig$-meson couples like a (pseudo) Goldstone of 
spontaneously broken symmetry.  
This is remarkable also considering  that the data-interpretation relies on the prediction of the gluon-fraction $\xx_g$.

\subsection{{The dilaton interpretation of the $D(0) < 0$ hypothesis}}

We turn to the $q^2 =0$ limit of the $D$-form factor, which has attracted considerable attention. 
Among the three nucleon form factors, the infrared interpretation due to mass and spin implies $A^N(0)=1$ and $J^N(0)=\tfrac{1}{2}$, while the value of $D^N(0)$ remains unclear.
In the Breit-frame, the $D$-form factor is associated with the spatial components of the energy-momentum tensor, 
which is interpreted as the internal force \cite{Polyakov:2002yz} associated with pressure, 
and it is argued   that for  
 a positive core pressure and a negative exterior pressure, one obtains $D(0)<0$  
 on grounds of mechanical stability \cite{Polyakov:2002yz,Hudson:2016gnq}.
This  conclusion and its interpretation  remains   actively debated \cite{Ji:2025gsq,Ji:2025qax,Lorce:2025oot}.\footnote{For the $\De$ this interpretation complicates as higher multipoles come into play \cite{Kim:2020lrs}.} 
 
We previously argued \cite{Stegeman:2025sca}  that the dilaton provides an alternative, microscopic interpretation 
which we revisit in the light of the higher spin cases.
Specifically, for any (light) hadron $H$, $D^H(0) < 0$ is guaranteed provided the $\sig$ dominates over the background 
\begin{equation}
\label{eq:LH}
D^H(0) = - \frac{r_\sig^H}{m_\sig^2} + b^H < 0 \;,
\end{equation} 
 since $\frac{r_\sig^H}{m_\sig^2} > 0$ must hold to approximately satisfy the conformal Ward identity. 
For light hadrons, this is supported by our fits to the gluon $D$-form factor in this work and the full form factors 
of the $m_\pi \approx 170\MeV$ data \cite{Stegeman:2025sca}.

Does $D(0) < 0$  hold in all systems? Not for the hydrogen atom, for which $D(0)>0$ has been found \cite{Ji:2022exr,Czarnecki:2023yqd,Freese:2024rkr}.
As the hydrogen atom has no dilaton, this does not contradict our interpretation, and in the case of the pressure interpretation it has been argued that long-range forces can provide exceptions to the rule \cite{Lorce:2025oot}.
Beyond this, we are not aware of any other firmly established counterexamples.

Let us be more precise. Unlike for the $N$ and the $\pi$, the situation for the $\rho$ and the $\De$ is far from clear.
For $D^\rho(0)$, the literature reports various possibilities: a negative sign from the Nambu--Jona-Lasino model computation \cite{Freese:2019bhb}, a positive sign from the light-cone sum rule computation \cite{Aliev:2020aih}, and essentially zero from a constituent quark model linked to generalised parton distribution functions \cite{Sun:2020wfo}.
For $D^\De(0)$ (and its $SU_F(3)$-related $D^\Omega(0)$), one finds a negative sign from light-cone sum rules \cite{Dehghan:2023ytx,Dehghan:2025eov}, whereas positive signs have been reported in a covariant quark-diquark approach \cite{Fu:2022rkn}.
Given this broad range of results, it is difficult to draw firm conclusions.
We therefore maintain  our interpretation of $D(0)<0$ for light hadrons based on our fits to the lattice data.

 \section{Heavy Hadron $D$-Form Factors are Different}
 \label{sec:eta}
 
Beyond  light hadrons, it is instructive to  examine heavy hadrons in the context of the 
dilaton hypothesis.  
In particular, the $\eta_{b,c}$-mesons, the beauty and charm analogues of positronium, are approximately stable since the width-to-mass ratios are small,
$(\Ga/m)|_{\eta_{b,c}} \approx 10^{-2}$~\cite{ParticleDataGroup:2024cfk}.  
Their gravitational form factors were recently computed in the Dyson-Schwinger framework using a contact interaction~\cite{Sultan:2024hep}.\footnote{
In that work, the $\pi$- and $\eta_s$-mesons are also shown.
The $\pi$-result qualitatively (if not quantitatively) agrees with the $m_\pi \approx 170~\MeV$ lattice data~\cite{Hackett:2023nkr}. For the $\eta_s$, which mixes into $\eta$--$\eta'$ in the real world, we are not aware of results 
one could compare to.}
From their \FIG 5 it is evident that the $\eta_{b,c}$ results do not exhibit a pronounced $\sig$-pole behaviour, certainly not of the magnitude  
$r_\sig^{\eta_{b,c}} \approx \tfrac{4}{3} m_{\eta_{b,c}}^2$. Stated differently, if we naively use the approximate 
light-hadron formula \eqref{eq:LH} neglecting the background, 
\begin{equation}
D^{\eta_{b[c]}}(0)\big|_{\text{naive}} = -  \frac{4}{3} \frac{ m_{\eta_{b[c]}}^2}{ m_\sig^2}  \quad 
\Rightarrow  \quad D^{\eta_{b}}(0)|_{\text{naive}} \approx -1500 \;, \quad D^{\eta_{c}}(0)|_{\text{naive}} \approx -50 \;,
\end{equation}
then it fails by orders of magnitude  compared to the results 
$D^{\eta_{(b,c)}}(0) \approx -(0.5,0.6) $  \cite{Sultan:2024hep} and  $-(0.13,0.28)$ \cite{Dwibedi:2026ozl}, and also 
the rather larger $-(-,4.7)$ {\cite{Xu:2024hfx}}.
A moment’s thought makes the strong suppression unsurprising since
$m_{\eta_{b,c}} = 2 m_{b,c}(1 + \ORD(\al_s^2))$,
so their masses are generated primarily by  explicit quark-mass terms.
Consequently, the dilaton need not act as a compensator to restore conformal symmetry 
 for these states. Stated differently, the explicit breaking alters the conformal Ward identities. 
It is therefore encouraging that the $\eta_{b,c}$-mesons are consistent with the behaviour anticipated within the dilaton interpretation.

\section{Conclusions and Discussion}
\label{sec:conc}

We investigated  gluon gravitational form factors of the 
$\pi$, $N$, $\rho$ and $\Delta$ using lattice QCD data at 
$m_\pi \approx 450$ and $m_\pi \approx 170\MeV$.
By fitting a simple $\sigma/f_0(500)$-pole supplemented by a background term 
(\ref{eq:Bansatz}, \ref{eq:Mansatz}), we tested the hypothesis that the $\sig$-meson acts as 
a (pseudo) dilaton; the (pseudo) Goldstone boson of spontaneously broken scale invariance. 
The extracted residues for all hadrons are compatible  within uncertainties
with the predictions of dilaton effective theory, see  \TAB\ref{tab:450} and \FIG\ref{fig:barchart}, 
supporting the hypothesis.
All residues are positive and, together with $\sig$-dominance, provide an alternative to the conventional $D$-term hypothesis in hadronic systems ($D(0) < 0$).

In addition, we find that the weak $\sig$-pole contributions in the $D^{\eta_{b,c}}$-form factor, as obtained in the Dyson-Schwinger framework, are consistent with the dilaton picture. 
Their masses are generated primarily by explicit breaking of scale symmetry. So, in this regime the dilaton is not required to couple to the mass term directly since the explicit breaking  alters the conformal Ward identities.

The overall agreement across hadrons of spin $j = 0, \tfrac{1}{2}, 1, \tfrac{3}{2}$ with the lattice data reinforce 
the universality of the dilaton coupling structure observed in our previous study~\cite{Stegeman:2025sca}.
While a  numerical coincidence cannot be ruled out, the consistent pattern across distinct hadronic channels strongly suggests a common origin in the spontaneous breaking of scale symmetry.
In this sense, the $\sig$ emerges as a \emph{pseudo dilaton}, a massive remnant 
of the potential Goldstone boson associated with broken scale invariance.

The behaviour of the $\sig$, particularly its complex pole for massless quarks, remains an open question.
We hope that future lattice QCD studies~\cite{Briceno:2016mjc,Briceno:2017qmb,Rodas:2023twk}, 
the analytic $S$-matrix bootstrap~\cite{Kruczenski:2022lot}  or dispersive methods  \cite{Caprini:2005zr}
will shed further light on this issue.

\paragraph{Acknowledgements:}
RS and RZ are supported by the STFC via the consolidated grants ST/T000600/1 and ST/X000494/1.
We are grateful to   Luigi Del Debbio, 
Nils Hermansson-Truedsson, Martin Hoferichter, Tej Kanwar,   C\'edric Lorc\'e, 
Keh-Fei  Liu,   Alex Ochirov, Maurizio Piai, Mannque Rho, Jacobo Ruiz de Elvira,  Cheryl Patrick, David Schaich and Raju Venugopalan   for discussions. 
A very special thanks goes to  the MIT-group members Daniel Hackett, Dimitra Pefkou and Phiala Shanahan for providing us with their form factor data, and especially Dimitra Pefkou for thorough correspondence.  

\appendix

\section{Gluon Fractions for $m_\pi \approx 170 \MeV$ and $\sig$-Mass Variations}
\label{app:additional}

\begin{figure}[h]
  \centering
    \includegraphics[width=0.49\linewidth]{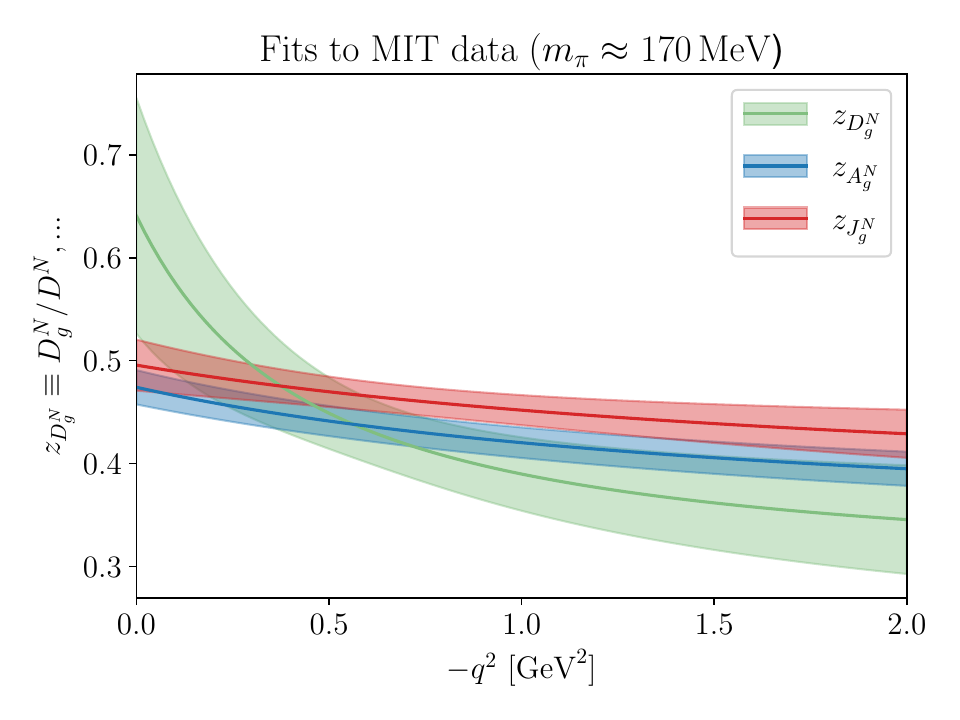}
      \includegraphics[width=0.49\linewidth]{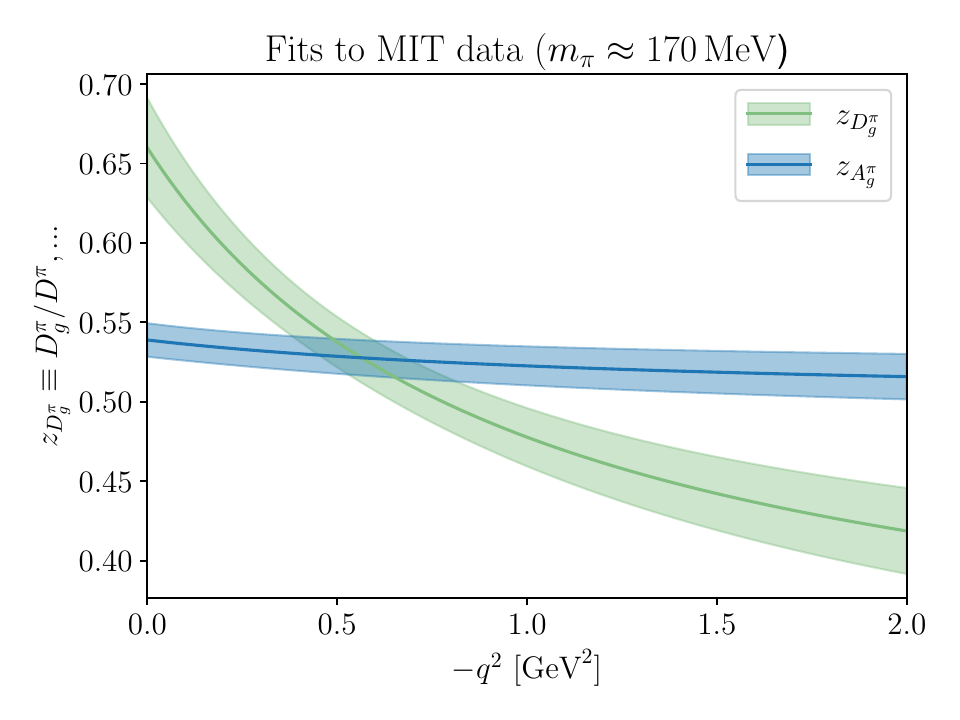}
         \caption{\small Ratio of gluon- to quark-part of the nucleon~\cite{Hackett:2023nkr} (left) and the pion 
         \cite{Hackett:2023rif} (right)  form factors of the MIT-data  at $m_\pi \approx 170\MeV$, 
     for $\mu = 2 \GeV$ in the $\MSbar$-scheme.
     We note that $A$ and $J$ are more or less constant  and roughly equal whereas the $D$-form factor raises in the 
     gluon-part towards zero momentum transfer.  
     }
  \label{fig:ratio}
\end{figure}
In this appendix we provide supplementary information supporting the main analysis.  \TAB\ref{tab:massvariation2}  shows the sensitivity of fitted residues to the assumed $\sig$-mass \eqref{eq:850} 
which enter 
the error estimate in  \TAB\ref{tab:450}.
 Figure \ref{fig:ratio} displays our fit   of the glue-fraction of the  $D$-form factor to the $m_\pi \approx 170\MeV$ data, 
 supporting the $\xx_g$-estimate in \EQ\eqref{eq:xgN}.

\begin{table}[h]
  \centering
  \small
  \setlength{\tabcolsep}{6pt} 
  \renewcommand{\arraystretch}{1.1} 
\begin{tabular}{l | r r r r || l |  rr }
  $m_\sigma$ & $\pi(450)$ & $N(450)$ & $\rho(450)$ & $\Delta(450)$ & $m_\sigma$  &  $\pi(170)$ & $N(170)$ \\ \hline
  $800\MeV$  & 0.48(26) & 1.65(63) & 0.17(31) & 1.55(83) &  $500\MeV$ & 0.48(12) & 0.66(11)\\
  \rowcolor{gray!15}   $850\MeV$  & 0.51(28) & 2.00(74) & 0.22(37) & 1.91(98) &  $550\MeV$ & 0.50(13) & 0.77(13)  \\
  $900\MeV$  & 0.54(29) & 2.41(87) & 0.28(44) & 2.3(1.2) &  $600\MeV$ & 0.52(13) & 0.91(15) \\
\end{tabular}
  \caption{\small The $\rsig{\sig,g}$-residues  fitted as in \TAB\ref{tab:450} for three different fixed $\sig$-masses 
  in order to assess the model-dependence of the fits. The values at $m_\sig=850\MeV$ are those already presented in \TAB\ref{tab:450} and provided here for comparison.}
  \label{tab:massvariation2}
\end{table}

\bibliographystyle{utphys}
\bibliography{../Dil-refs.bib}

\end{document}